\documentclass[twocolumn]{aastex61}
\usepackage{color}

\newcommand\aastex{AAS\TeX}

\def\star{\hbox{LTT~1445A}} 
\def\system{\hbox{LTT~1445ABC}} 
\def\planet{\hbox{LTT~1445Ab}} 
\def\mdot{\hbox{M$_{\odot}$}} 
\def\rearth{\hbox{R$_{\oplus}$}} 
\def\mearth{\hbox{M$_{\oplus}$}} 
\def\tess{\hbox{$TESS$}} 
\def\kepler{\hbox{$Kepler$}} 
\def\hst{\hbox{$HST$}}
\def\pers{\hbox{s$^{-1}$}}

\def\rplanet{\hbox{$1.38^{+0.13}_{-0.12}$}} 
\def\muplanet{\hbox{$8.4$}} 
\def\mplanet{\hbox{$2.2^{+1.7}_{-2.1}$}} 
\def\pplanet{\hbox{$5.35882^{+0.00030}_{-0.00031}$}} 
\def\teqplanet{\hbox{$433^{+28}_{-27}$}}

\providecommand{\bjdtdb}{\ensuremath{\rm {BJD_{TDB}}}}

\providecommand{\msun}{\ensuremath{\,M_\Sun}}
\providecommand{\rsun}{\ensuremath{\,R_\Sun}}

\shorttitle{\aastex\ LTT 1445ABC}
\shortauthors{Winters et al.}

\begin{document}

\title{Three Red Suns in the Sky: A Transiting, Terrestrial Planet in a Triple M Dwarf System at 6.9 Parsecs}

\correspondingauthor{Jennifer G. Winters}
\email{jennifer.winters@cfa.harvard.edu}

\author[0000-0001-6031-9513]{Jennifer G.\ Winters}
\affil{Center for Astrophysics $\vert$ Harvard \& Smithsonian, 60 Garden Street, Cambridge, MA 02138, USA}

\author[0000-0001-8726-3134]{Amber A.\ Medina}
\affil{Center for Astrophysics $\vert$ Harvard \& Smithsonian, 60 Garden Street, Cambridge, MA 02138, USA}

\author{Jonathan M.\ Irwin}
\affil{Center for Astrophysics $\vert$ Harvard \& Smithsonian, 60 Garden Street, Cambridge, MA 02138, USA}

\author[0000-0002-9003-484X]{David Charbonneau}
\affil{Center for Astrophysics $\vert$ Harvard \& Smithsonian, 60 Garden Street, Cambridge, MA 02138, USA}

\author[0000-0002-8462-515X]{Nicola Astudillo-Defru}
\affil{Departamento de Matem\'atica y F\'isica Aplicadas, Universidad Cat\'olica de la Sant\'isima Concepci\'on, Alonso de Rivera 2850, Concepci\'on, Chile}

\author[0000-0003-2159-1463]{Elliott P.\ Horch}
\affil{Department of Physics, Southern Connecticut State University, 501 Crescent Street, New Haven, CT 06515, USA}

\author[0000-0003-3773-5142]{Jason D.\ Eastman}
\affil{Center for Astrophysics $\vert$ Harvard \& Smithsonian, 60 Garden Street, Cambridge, MA 02138, USA}

\author[0000-0002-1864-6120]{Eliot Halley Vrijmoet}
\affil{Department of Physics and Astronomy, Georgia State University, Atlanta, GA 30302-4106, USA}

\author{Todd J.\ Henry}
\affil{RECONS Institute, Chambersburg, Pennsylvania, 17201, USA}

\author[0000-0001-8274-6639]{Hannah Diamond-Lowe}
\affil{Center for Astrophysics $\vert$ Harvard \& Smithsonian, 60 Garden Street, Cambridge, MA 02138, USA}

\author[0000-0001-9065-6633]{Elaine Winston}
\affil{Center for Astrophysics $\vert$ Harvard \& Smithsonian, 60 Garden Street, Cambridge, MA 02138, USA}

\author[0000-0001-7139-2724]{Thomas Barclay}
\affil{NASA Goddard Space Flight Center, 8800 Greenbelt Road, Greenbelt, MD 20771, USA}
\affil{University of Maryland, Baltimore County, 1000 Hilltop Circle, Baltimore, MD 21250, USA}

\author{Xavier Bonfils}
\affil{Universit\'e Grenoble Alpes, CNRS, IPAG, F-38000 Grenoble, France}

\author{George R.\ Ricker} 
\affil{Department of Physics and Kavli Institute for Astrophysics and Space Science, Massachusetts Institute of Technology, Cambridge, MA 02139, USA}

\author[0000-0001-6763-6562]{Roland Vanderspek}
\affil{Department of Physics and Kavli Institute for Astrophysics and Space Science, Massachusetts Institute of Technology, Cambridge, MA 02139, USA}

\author[0000-0001-9911-7388]{David W.\ Latham}
\affil{Center for Astrophysics $\vert$ Harvard \& Smithsonian, 60 Garden Street, Cambridge, MA 02138, USA}

\author[0000-0002-6892-6948]{Sara Seager}
 \affiliation{Department of Physics and Kavli Institute for Astrophysics and Space Research, Massachusetts Institute of Technology, Cambridge, MA 02139, USA}
\affiliation{Department of Earth, Atmospheric and Planetary Sciences, Massachusetts Institute of Technology, Cambridge, MA 02139, USA}
\affiliation{Department of Aeronautics and Astronautics, MIT, 77 Massachusetts Avenue, Cambridge, MA 02139, USA}

\author[0000-0002-4265-047X]{Joshua N.\ Winn}
\affil{Department of Astrophysical Sciences, Princeton University, Princeton, NJ 08544, USA}

\author[0000-0002-4715-9460]{Jon M.\ Jenkins}
\affil{NASA Ames Research Center, Moffett Field, CA 94035, USA}

\author{St\'ephane Udry}
\affil{Observatoire de Gen\`eve, Universit\'e de Gen\`eve, 51 ch. des Maillettes, 1290 Sauverny, Switzerland}

\author[0000-0002-6778-7552]{Joseph D.\ Twicken}
\affil{SETI Institute, Moffett Field, CA 94035, USA}
\affil{NASA Ames Research Center, Moffett Field, CA 94035, USA}

\author{Johanna K.\ Teske}
\affil{Observatories of the Carnegie Institution for Science, 813 Santa Barbara Street, Pasadena, CA 91101, USA}
\affil{Hubble Fellow}

\author{Peter Tenenbaum}
\affil{SETI Institute, Moffett Field, CA 94035, USA}
\affil{NASA Ames Research Center, Moffett Field, CA 94035, USA}

\author{Francesco Pepe}
\affil{Observatoire de Gen\`eve, Universit\'e de Gen\`eve, 51 ch. des Maillettes, 1290 Sauverny, Switzerland}

\author[0000-0001-9087-1245]{Felipe Murgas}
\affil{Instituo de Astrof\'isica da Canarias (IAC), 38205 La Laguna, Tenerife, Spain}
\affil{Departamento de Astrof\'isica, Universidad de La Laguna (ULL), E-38206 La Laguna, Tenerife, Spain} 

\author[0000-0002-0638-8822]{Philip S.\ Muirhead}
\affil{Department of Astronomy \& The Institute for Astrophysical Research, Boston University, 725 Commonwealth Avenue, Boston, MA 02215, USA}

\author{Jessica Mink}
\affil{Center for Astrophysics $\vert$ Harvard \& Smithsonian, 60 Garden Street, Cambridge, MA 02138, USA}

\author{Christophe Lovis}
\affil{Observatoire de Gen\`eve, Universit\'e de Gen\`eve, 51 ch. des Maillettes, 1290 Sauverny, Switzerland}

\author[0000-0001-8172-0453]{Alan M.\ Levine}
\affil{Department of Physics and Kavli Institute for Astrophysics and Space Science, Massachusetts Institute of Technology, Cambridge, MA 02139, USA}

\author[0000-0002-2437-2947]{S\'ebastien L\'epine}
\affil{Department of Physics and Astronomy, Georgia State University, Atlanta, GA 30302-4106, USA}

\author[0000-0003-0193-2187]{Wei-Chun Jao}
\affil{Department of Physics and Astronomy, Georgia State University, Atlanta, GA 30302-4106, USA}

\author{Christopher E.\ Henze}
\affil{NASA Ames Research Center, Moffett Field, CA 94035, USA}

\author{G\'abor Fur\'esz}
\affil{Department of Physics and Kavli Institute for Astrophysics and Space Science, Massachusetts Institute of Technology, Cambridge, MA 02139, USA}

\author[0000-0003-0536-4607]{Thierry Forveille}
\affil{Universit\'e Grenoble Alpes, CNRS, IPAG, F-38000 Grenoble, France}

\author{Pedro Figueira}
\affil{European Southern Observatory, Alonso de C\'ordova 3107, Vitacura, Regi\'on Metropolitana, Chile}
\affil{Instituto de Astrof\'isica e Ci\^{e}ncias do Espa\c{c}o, Universidade do Porto, CAUP, Rua das Estrelas, 4150-762 Porto, Portugal}


\author[0000-0002-9789-5474]{Gilbert A.\ Esquerdo}
\affil{Center for Astrophysics $\vert$ Harvard \& Smithsonian, 60 Garden Street, Cambridge, MA 02138, USA}

\author[0000-0001-8189-0233]{Courtney D.\ Dressing}
\affil{Astronomy Department, University of California, Berkeley, CA 94720, USA}

\author{Rodrigo F.\ D\'iaz}
\affil{Universidad de Buenos Aires, Facultad de Ciencias Exactas y Naturales, Buenos Aires, Argentina}
\affil{CONICET - Universidad de Buenos Aires, Instituto de Astronom\'ia y F\'isica del Espacio (IAFE), Buenos Aires, Argentina}

\author{Xavier Delfosse}
\affil{Universit\'e Grenoble Alpes, CNRS, IPAG, F-38000 Grenoble, France}

\author[0000-0002-7754-9486]{Christopher~J.~Burke}
\affil{Department of Physics and Kavli Institute for Astrophysics and Space Science, Massachusetts Institute of Technology, Cambridge, MA 02139, USA}

\author{Fran\c{c}ois Bouchy}
\affil{Observatoire de Gen\`eve, Universit\'e de Gen\`eve, 51 ch. des Maillettes, 1290 Sauverny, Switzerland}

\author{Perry Berlind}
\affil{Center for Astrophysics $\vert$ Harvard \& Smithsonian, 60 Garden Street, Cambridge, MA 02138, USA}

\author[0000-0003-3208-9815]{Jose-Manuel Almenara}
\affil{Universit\'e Grenoble Alpes, CNRS, IPAG, F-38000 Grenoble, France}

\begin{abstract}

We present the discovery from \tess ~data of LTT~1445Ab. At a distance of 6.9 parsecs, it is the second nearest transiting exoplanet system found to-date, and the closest one known for which the primary is an M dwarf. The host stellar system consists of three mid-to-late M dwarfs in a hierarchical configuration, which are blended in one \tess ~pixel. We use MEarth data and results from the SPOC DV report to determine that the planet transits the primary star in the system. The planet has a radius \rplanet ~\rearth, an orbital period of \pplanet ~days, and an equilibrium temperature of \teqplanet ~K. With radial velocities from HARPS, we place a three-sigma upper mass limit of \muplanet ~\mearth ~on the planet. LTT~1445Ab provides one of the best opportunities to-date for the spectroscopic study of the atmosphere of a terrestrial world. 
We also present a detailed characterization of the host stellar system. We use high-resolution spectroscopy and imaging to rule out the presence of any other close stellar or brown dwarf companions. Nineteen years of photometric monitoring of A and BC indicates a moderate amount of variability, in agreement with that observed in the \tess ~light curve data. We derive a preliminary astrometric orbit for the BC pair that reveals an edge-on and eccentric configuration. The presence of a transiting planet in this system hints that the entire system may be co-planar, implying that the system may have formed from the early fragmentation of an individual protostellar core.

\end{abstract}


\section{Introduction} \label{sec:intro}







Until the advent of large space missions capable of spatially resolving rocky planets from their host stars, the only terrestrial exoplanets that will be spectroscopically accessible will be those that orbit nearby, mid-to-late M dwarfs \citep{NASESS(2018)}. Transiting examples of such planets are particularly advantageous, as they allow the unambiguous determination of masses, radii, mean densities and surface gravities, and permit their atmospheres to be probed with the technique of transmission spectroscopy. Yet, even with the large apertures of upcoming facilities, such studies will be photon starved: it may be possible to search for molecular oxygen in the atmospheres of terrestrial exoplanets with the upcoming cohort of ground-based giant, segmented mirror telescopes (GSMTs), but studies indicate that even marginal detections will be feasible only for stars within 15~parsecs and no larger than 0.3~R$_{\Sun}$ \citep{Snellen(2013),Rodler(2014),LopezMorales(2019)}. The eagerly awaited {\it James Webb Space Telescope} ({\it JWST}) may also be able to detect key molecules such as water, methane, and carbon dioxide in the atmospheres of terrestrial exoplanets, but again demands parent stars that are similarly nearby, and small \citep{Morley(2017)}.

Within 15~parsecs, there are 411 M dwarfs with masses between 0.3 and 0.1 \mdot, and perhaps an additional 60 systems between 0.1 M$_\Sun$ and the main-sequence cut-off \citep{Winters(2018AAS),Winters(2019a),Gagliuffi(2019)}. How many transiting terrestrial worlds might we expect within this sample of stars? \citet{Dressing(2015)} analyzed the data from the {\it Kepler} mission and found that, on average, M dwarfs host 2.5 planets smaller than 4~R$_\Earth$ with periods less than 200~days. Considering only planets with radii between $1.0-1.5$~R$_\Earth$ and periods less than 50~days, they found a mean number of planets per M dwarf of 0.56. Importantly, these stellar primaries were typically early M dwarfs, roughly twice as massive as the mid-to-late M dwarfs required to enable the atmospheric studies described above. Although efforts are underway to use K2 data to determine the rate of planet occurrence for the less massive M dwarfs \citep[e.g.,][]{Dressing(2019)}, it is currently an open question whether they host small planets with the same frequency as their more massive counterparts. 

For stars less massive than 0.3~M$_\Sun$ and within 15~parsecs, four families of transiting, terrestrial planets are known: GJ~1132bc \citep{Berta-Thompson(2015),Bonfils(2018)}; LHS~1140bc \citep{Dittmann(2017a),Ment(2019)}; TRAPPIST-1bcdefgh \citep{Gillon(2016),Gillon(2017),Grimm(2018)}; and LHS~3844b \citep{Vanderspek(2019)}. Yet, the closest of these lies at 12~parsecs, for which {\it JWST} and the GSMTs may still be at pains to access. Thus, there is great interest within the community to identify even closer examples of such systems. 


The $Transiting ~Exoplanet ~Survey ~Satellite$ \citep[\tess; ][]{Ricker(2015)} mission is now one year into its 2-year prime mission to scan most of the sky in search of the transiting planets that are most amenable to follow-up study. 
We report here the detection with \tess ~data of the second closest known transiting exoplanet system, \system ~(TIC~98796344, TOI~455), and the nearest one for which a terrestrial planet transits a low-mass star. The planet is 6.9~parsecs away, and orbits one member of a stellar triplet. Multi-star systems present numerous challenges that sometimes deter planet hunters: astrometric perturbations from stellar companions at small separations can hinder the measurement of the trigonometric parallax of the system; the presence of bound companions can result in trends in the radial velocities of a star that can mask the signals of planets; and, light contamination from close stellar companions in the photometry of a host star can result in an underestimated planet radius \citep{Ciardi(2015),Furlan(2017),Hirsch(2017)}. Yet, these complications are also opportunities to measure the stellar orbits and investigate the potential formation scenarios for the planets that are found within; indeed all of these features are present in the system that is the subject of our study. We present here the discovery of the planet and a description of the host star system. We first provide a detailed portrait of the host star system in \S \ref{sec:host_system}. We then detail the observations in \S \ref{sec:data}. In \S \ref{sec:analysis}, we present our analysis of the data. Finally, in \S \ref{sec:disc} we discuss the implications of this planet and the opportunity it presents for characterization of its atmosphere. 

\section{Description of the Host Stellar System}\label{sec:host_system}

\begin{figure*}
\centering
\includegraphics[scale=.70,angle=0]{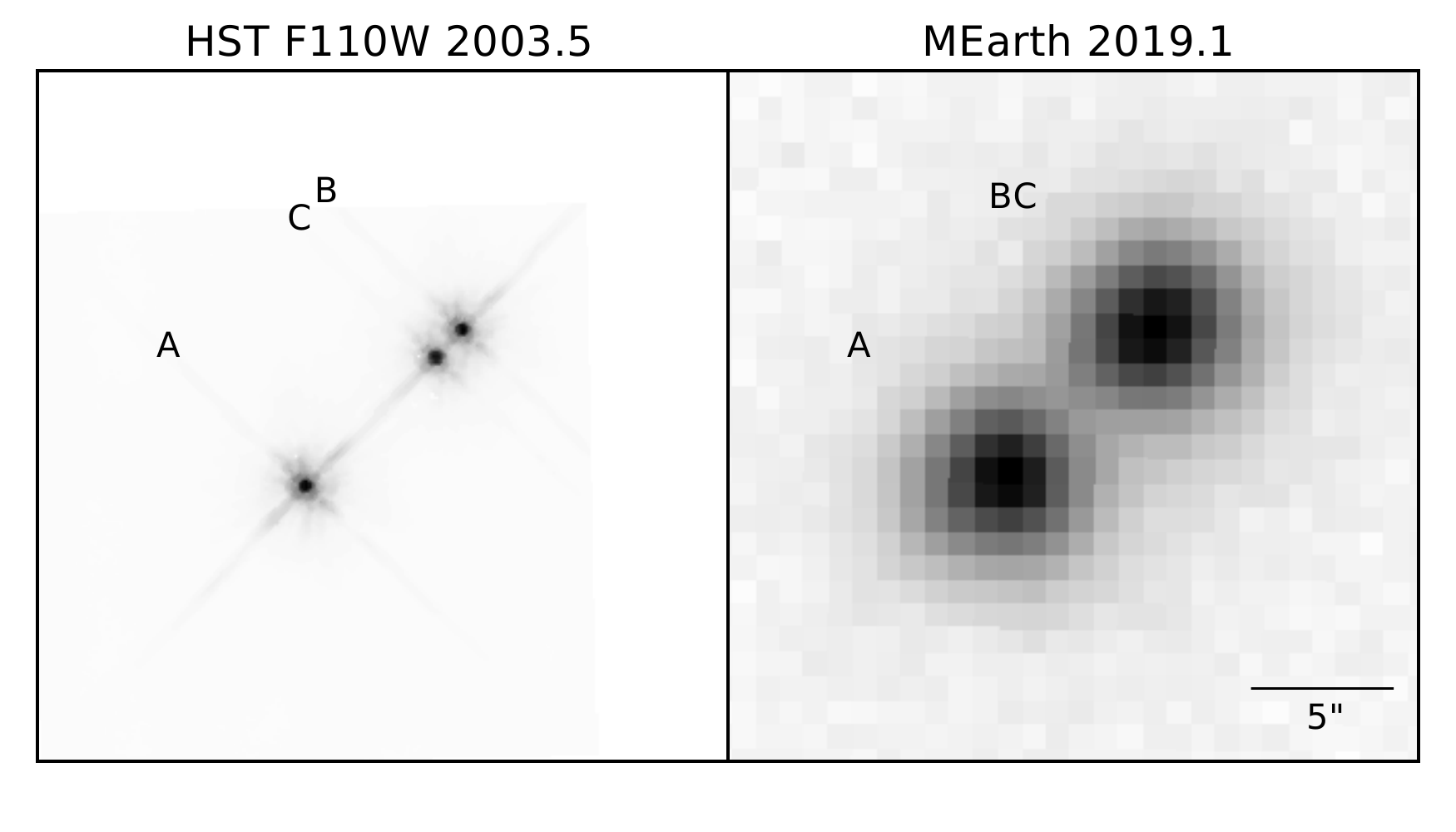}
\caption{LTT~1445. Left: \hst ~NICMOS image in the F110W filter taken in 2003; Right: MEarth image taken in 2019. North is up, East is left. The BC pair are blended in the ground-based MEarth image. We note that all three components fall on one 21\arcsec ~square \tess ~pixel. \label{fig:finders}}
\end{figure*}

The host system, LTT~1445ABC \citep{Luyten(1957),Luyten(1980a)}, is a nearby, hierarchical trio of mid-to-late M dwarfs. \citet{Rossiter(1955)} is the first observer to have noted relative astrometry for LTT~1445ABC using visual micrometry. In two observations made near the beginning of 1944 (specifically, Besselian years 1943.960 and 1944.027), the primary star was measured to have a separation from the B component of 3\farcs03 and 3\farcs51. In those same observations, the BC subsystem was measured to have a separation of approximately 1\farcs3. Since then, the separation of the primary relative to the subsystem has increased to a maximum value of 7\farcs706 in 2003 \citep{Dieterich(2012)}, and is now apparently decreasing, with the most recent value of 7\farcs10 obtained in 2017, according to data available in the Washington Double Star (WDS) Catalog\footnote{\url{https://www.usno.navy.mil/USNO/astrometry/optical-IR-prod/wds/WDS}} \citep{Mason(2009b)}. In contrast, for much of the time since 1944, the BC subsystem has been on a trajectory of decreasing separation; however, the most recent speckle observations appear to show that this trend has now
reversed, and the separation is growing larger. As shown in Figure \ref{fig:finders}, the three components are visible in an archival $Hubble ~Space ~Telescope$ ~($HST$) NICMOS image (left panel), but the B and C components are blended in our ground-based image from MEarth-South (right panel). 



As reported by \citet{Henry(2018)} with over eighteen years of RECONS\footnote{REsearch Consortium On Nearby Stars; \url{www.recons.org}.} astrometry data, the position of the primary star exhibits an astrometric perturbation due to the presence of the BC pair. We describe a preliminary orbit for the BC pair below in \S \ref{subsec:bc_orbit}. 




Because of the nearly equal brightnesses of the A and blended BC components of the system, there has been confusion in the literature regarding their designations. Convention dictates that the primary component is the brightest in the $V-$band and is therefore the most massive of all stars in a multiple system\footnote{A rare exception is the case of a red dwarf--white dwarf pair where the two stars may have equal fluxes or equal masses, but not both at a given wavelength.}. The confusion appears to have originated with Luyten, who may not have been aware of Rossiter's work on this system. The Luyten Two Tenths (LTT) Catalogue \citep{Luyten(1957)} lists only one entry for the system, but the New Luyten Two Tenths (NLTT) Catalogue \citep{Luyten(1980a)} lists two entries with identical coordinates and proper motions. The brighter star (m$_R$ $=$ 11.1 mag, m$_{pg}$ $=$ 12.7 mag) is noted as LP~771-95; the fainter star (m$_R$ $=$ 11.8 mag, m$_{pg}$ $=$ 13.5 mag) is noted as LP~771-96. The entry for LP~771-96 includes the note, `Comp. to 95, 121\arcdeg, 4\farcs5', which implies that the secondary is the southeast component. The WDS entry for this result has been edited so that the position angle between the primary and secondary is 301\arcdeg, presumably to bring it into agreement with results from Rossiter who had already resolved the brighter component into a stellar pair. For the remainder of this paper, we refer to the stellar system as LTT~1445ABC, where A is the southeast component and BC is the blended northwest component. 


While LTT~1445A\footnote{Other names: TIC~98796344, TOI~455, L~730-18, BD-17~588A, RST~2292A, WDS~J03019-1633A,  2MASS~J03015142-1635356, Gaia~DR2~5153091836072107136.} is brighter than LTT~1445BC in the $V$ filter, the opposite is true in the $K$ filter. Thus, if one were unaware of the binarity of the B component, mass estimates calculated using a mass-luminosity relation (MLR) will be discrepant, depending which filter relation is used: the A component will be more massive with a $V-$band MLR calculation, while the B component will be more massive with a $K-$band MLR calculation. However, we know that LTT~1445BC\footnote{Other names: TIC~98796342, BD-17~588B, RST~2292BC, WDS~J03019-1633B, 2MASS~J03015107-1635306, Gaia~DR2~5153091836072107008.} is a sub-arcsecond binary with both optical and infrared delta-magnitudes ($\Delta$mag) reported in the literature. Once the photometry of the BC components is deblended into their individual photometric magnitudes, their overluminosity in the infrared is resolved and the A component is the brightest and most massive star in the system in both the optical and infrared filters. 

We calculated the \tess ~magnitudes, $T$, for all three components from ($I_{\rm KC} - K_{\rm s}$) colors\footnote{Henceforth, we omit the subscripts on these filters. The central wavelengths are 8075 \AA ~and 2.159 $\mu$m ~for the $I$ and $K$ filters, respectively.} using relations appropriate for M dwarfs developed by Guillermo Torres (private communication). The transformation is valid for M dwarfs with near-solar metallicity (-1.0 $\leq$ [Fe/H] $\leq$ 0.5) and has a residual scatter of 0.013 mag. We provide the relation here:

\vspace{0.3cm}

$T = I - 1.2457 + 1.54056*(I-K) - 0.49790*(I-K)^2 + 0.04539*(I-K)^3$

\vspace{0.3cm}

The transformation from ($I-K$) to $T$ for the primary star was straightforward. For the secondary pair, we first deblended the $I-$ and $K-$band magnitudes using $\Delta$mags reported in the literature: $\Delta$$I$ of 0.66$\pm$0.07 mag by \citet{Henry(2006)} and $\Delta$$K$ of 0.52$\pm$0.03 mag, which we calculated by averaging the values of $\Delta$F207M and $\Delta$F222M by \citet{Dieterich(2012)}. We also deblended the $V-$ and $R-$ band photometry. We list the stellar system parameters in Table \ref{tab:system_info}.


\begin{deluxetable*}{lccccc}
\tablecaption{System Parameters for LTT~1445ABC \label{tab:system_info}}
\tablecolumns{6}
\tablenum{1}
\tablewidth{0pt}
\tablehead{
\colhead{Parameter}  &
\colhead{A}      &
\colhead{BC}     &
\colhead{B - Deblended}     &
\colhead{C - Deblended}     &
\colhead{Reference}      
} 
\startdata
RA (2000.0) (hh:mm:ss)                     & 03:01:51.39         & 03:01:51.04       &   \nodata         &  \nodata         & 3,3 \\
Decl. (2000.0) (dd:mm:ss)                  & $-$16:35:36.1       & $-$16:35:31.1     &   \nodata         &  \nodata         & 3,3 \\
Proper Motion Mag. (mas yr$^{-1}$)         & 456.5$\pm$0.2       & 479.4$\pm$0.3     &   \nodata         &  \nodata         & 2,3 \\
Proper Motion PA (deg)                     & 234.0$\pm$0.07      & 234.1$\pm$0.08    &   \nodata         &  \nodata         & 2,3 \\
Parallax (mas)                             & 145.55$\pm$0.08     & 142.57$\pm$2.03   &   \nodata         &  \nodata         & 2,3;3\\
$T$ (mag)                                  &  8.88$\pm$0.02      &   8.80$\pm$0.02J  &   9.27$\pm$0.07   &  9.92$\pm$0.07   &1,1,1,1 \\
$V_{\rm J}$ (mag)                          & 11.22$\pm$0.02      &  11.37$\pm$0.03J  &  11.78$\pm$0.09   &  12.64$\pm$0.09  &3,3,1,1 \\
$R_{\rm KC}$ (mag)                         & 10.07$\pm$0.02      &  10.13$\pm$0.02J  &  10.57$\pm$0.04   &  11.32$\pm$0.04  &3,3,1,1 \\
$I_{\rm KC}$ (mag)                         &  8.66$\pm$0.02      &   8.58$\pm$0.02J  &   9.05$\pm$0.07   &  9.71$\pm$0.07   &3,3,1,1 \\
$J$ (mag)                                  &  7.29$\pm$0.02      &   7.11$\pm$0.02J  &   \nodata         &  \nodata          & 4,4   \\
$H$ (mag)                                  &  6.77$\pm$0.04      &   6.56$\pm$0.02J  &   \nodata         &  \nodata          & 4,4 \\
$K_{\rm S}$ (mag)                          &  6.50$\pm$0.02      &   6.29$\pm$0.02J  &   6.81$\pm$0.04  &  7.33$\pm$±0.04    & 4,4 \\
Mass (\mdot)                               &  0.257$\pm$0.014    &   \nodata         &  0.215$\pm$0.014  &  0.161$\pm$0.014  & 1,1,1 \\
Radius (R$_{\odot}$)                       &  0.268$\pm$0.027    &   \nodata         &  0.236$\pm$0.027  &  0.197$\pm$0.027  & 1,1,1 \\
\enddata
\tablecomments{`J' indicates that the listed parameter is `joint' and includes both the B and C components.}
\tablerefs{
(1) this work;
(2) \citet{GaiaDR2(2018)};
(3) \citet{Henry(2018)};
(4) \citet{Skrutskie(2006)}.
}
\end{deluxetable*}



\subsection{Orbit Calculation of LTT~1445BC}
\label{subsec:bc_orbit}

We used the high-resolution astrometry of observations appearing in the Fourth Interferometric Catalog\footnote{\url{https://www.usno.navy.mil/USNO/astrometry/optical-IR-prod/wds/int4}} (FIC), plus the addition of previously unpublished data using the Differential Speckle Survey Instrument (DSSI) speckle camera \citep{Horch(2009)} at the WIYN telescope in 2012, to compute a preliminary visual orbit for the BC subsystem. Table \ref{tab:bc_obs} lists the observations used in the orbit calculation, including previously published DSSI observations of this pair \citep{Horch(2015b),Horch(2017)}. Only data from 2003 to the present were used; data before this did not use high-resolution techniques, and therefore the astrometry would generally be of lower precision. We used the method of \citet{MacKnight(2004)}, which takes as input an upper limit and a lower limit for each of the seven orbital elements and first computes a grid search to find the elements within the ranges selected that minimize the squared observed-minus-predicted differences in the secondary's position. After those elements are found, a second calculation is done to refine those orbital elements using the downhill simplex algorithm. Uncertainties in orbital elements are estimated by adding Gaussian random deviations of the expected astrometric uncertainty to all of the observed position angles and separations and recomputing the orbit many times. This yields a sample distribution for each orbital element; the uncertainty is calculated to be the standard deviation of the distribution in each case.  

\begin{figure}
\includegraphics[scale=.55,angle=0]{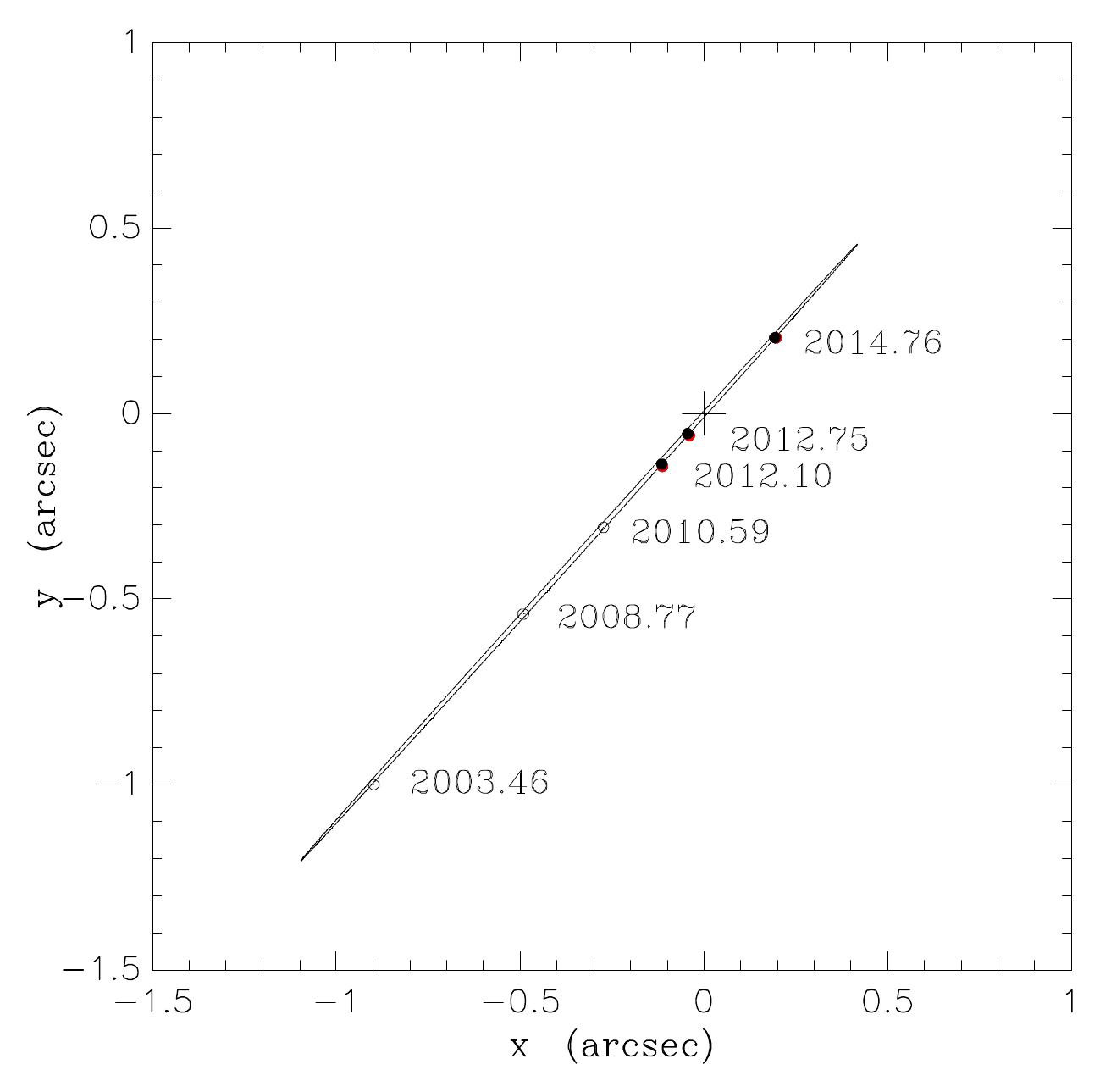}
\caption{Preliminary astrometric orbit of LTT~1445C relative to LTT~1445B. Data points appearing in the Fourth Interferometric Catalog are shown with open circles. The three pairs of observations obtained with DSSI are shown as filled circles, with red indicating the 692nm filter and black indicating the 880nm filter. The cross indicates the position of LTT~1445B. North is up; East is left. \label{fig:BC_orbit}}
\end{figure}

A reanalysis of the 2014 DSSI data indicates that the quadrant of the secondary in those images is ambiguous, a situation that can arise in speckle imaging when observing fainter targets. Using the position angle values shown in the FIC results in an orbit that is highly eccentric ($e=0.9$) and implies a mass sum for the BC subsystem of $0.63 \pm 0.28$ solar masses. On the other hand, if one reverses the quadrant of those observations by adding 180 degrees to the position angle, the derived parameters indicate that the pair has a somewhat eccentric, edge-on orbit with a period of roughly 36 years. While either orbit is possible, at this point we judge the the latter as more likely because the residuals to the orbit fit are significantly smaller. We list the orbital elements derived in that case in Table \ref{tab:bc_orbit}. Because the data span only 11 years (roughly one-third of the orbital period shown in Table 3), this results in large uncertainties in some of the orbital elements, particularly the semi-major axis and the period.  High-quality orbital elements are not likely to be obtained for another decade, when observations will hopefully be available to clearly show the orbital progress. We show the visual orbit we have calculated in Figure \ref{fig:BC_orbit}. We show the ephemeris predictions and residuals for all observations used in the orbit calculation in Table \ref{tab:bc_eph}.

\begin{deluxetable*}{llllrrccc}
\tabletypesize{\scriptsize}
\tablenum{2}
\tablecaption{Astrometry and Photometry for Observations Used in the Orbit Calculation for LTT~1445BC \label{tab:bc_obs}}
\tablehead{ 
\colhead{Date} & 
\colhead{$\theta$} & \colhead{$\rho$} & 
\colhead{$\Delta m$} & 
\colhead{$\lambda$} & 
\colhead{$\Delta \lambda$} &
\colhead{Tel. Dia.} &
\colhead{Technique} &
\colhead{Reference}\\
\colhead{} & 
\colhead{($^{\circ}$)} & 
\colhead{(${\prime \prime}$)} &
\colhead{(mag)} &
\colhead{(nm)} & 
\colhead{(nm)}  &
\colhead{(m)} &&
} 
\startdata 
2003.4620 & 138.1   &    1.344    &   0.56   &     1797  & 68  & 2.4  & $HST$ NICMOS  & 2 \\
2008.7675 & 137.7   &    0.7305   &   1.4    &      551  & 22  & 4.1  & Speckle     & 6 \\
2010.594  & 138.41  &    0.41     &   0.52   &     2150  & 320 & 3.0  & AO          & 5 \\
2012.0963 & 141.8   &    0.1812   &   2.16   &      692  & 40  & 3.5  & Speckle     & 4 \\
2012.0963 & 140.0   &    0.1777   &   1.80   &      880  & 50  & 3.5  & Speckle     & 4 \\
2012.7516 & 146.5   &    0.0710   &   1.05   &      692  & 40  & 3.5  & Speckle     & 1 \\
2012.7516 & 141.4   &    0.0694   &   0.87   &      880  & 50  & 3.5  & Speckle     & 1 \\
2014.7557 & 316.1\tablenotemark{*}   &    0.2838   &   1.47   &      692  & 40  & 4.3  & Speckle     & 3 \\
2014.7557 & 316.6\tablenotemark{*}   &    0.2824   &   1.03   &      880  & 50  & 4.3  & Speckle     & 3 \\
\enddata 
\tablenotetext{*}{Quadrant ambiguous; the position angle here has been changed by 180 degrees relative to the original result.}
\tablerefs{
(1) This paper;
(2) \citet{Dieterich(2012)};
(3) \citet{Horch(2015b)};
(4) \citet{Horch(2017)};
(5) \citet{Rodriguez(2015)};
(6) \citet{Tokovinin(2010a)}.
}
\end{deluxetable*}

Despite the very tentative nature of this orbit, the inclination is already very well constrained and the speckle data points have residuals that are fairly consistent with the known measurement precision of the technique; typical measurement uncertainties would be in the range of 2-3 mas in separation, and 0.5-1 degree in position angle at the telescope used, for example. Using the {\it Gaia} DR2 parallax reported for the primary star and the orbital results for period and semi-major axis, we derive a total mass for the BC subsystem of $0.39\pm0.09$ \mdot. The uncertainty in the mass is calculated from the independent trials of the orbit described above. Each trial gives a period and semi-major axis, and each is used, together with a randomly-chosen parallax deviate from a Gaussian distribution of the same mean of the Gaia parallax and standard deviation of the Gaia parallax uncertainty. The standard deviation of these mass values is our estimate of the mass uncertainty. This result is consistent with the sum of the masses estimated from the mass-$M_K$ relation \citep{Benedict(2016)} using the deblended $K-$band photometry (0.221$\pm$0.014 $+$ 0.165$\pm$0.014 \mdot $=$ 0.386$\pm$0.020 \mdot). Using the total mass of the three stellar components and an average angular separation of 5\arcsec ~(corresponding to 34 AU), we estimate the period of the A-BC orbit to be roughly 250 years. No further analysis of this orbit has been done.


Photometric information on the BC pair is difficult to interpret at this stage. \citet{Henry(2006)} noted a decrease in brightness of 0.3 magnitudes in a blended image taken in 1999. This was one of a series of observations taken as a part of the RECONS parallax effort, and those authors suggested that they had viewed a potential eclipsing event. The \citet{Tokovinin(2010a)} measure from 2008 notes the magnitude difference obtained is of reduced quality. In the recent sequence of speckle measures with the DSSI speckle instrument, including the measures presented here from October 2012, there are three measures of the magnitude difference in both 692nm and 880nm filters, but they have large scatter. Computing the average and standard error from these measurements, we obtain $\Delta692nm = 1.56 \pm 0.32$ and $\Delta880nm = 1.23 \pm 0.29$.  These values carry larger uncertainty than expected; more observations are warranted.



\begin{deluxetable}{lc}
\tabletypesize{\small}
\tablecaption{Preliminary Orbital Parameters for LTT~1445BC \label{tab:bc_orbit}}
\tablecolumns{2}
\tablenum{3}

\tablehead{
\colhead{Parameter} & 
\colhead{Value} 
}
\startdata
Orbital Period (years)	        & 36.2$\pm$5.3	\\
Semi-major axis (arcseconds)	& 1.159$\pm$0.076 \\
Inclination (degrees)        	& 89.64$\pm$0.13  \\
$\Omega$ (degrees)             	& 137.63$\pm$0.19 \\
T$_0$ (Besselian year)        	& 2019.2$\pm$1.7 \\
Eccentricity                	& 0.50$\pm$0.11	 \\
$\omega$ (degrees)              	& 209$\pm$13 \\
\enddata
\end{deluxetable}

\begin{deluxetable}{lrrrr}
\tablewidth{0pt}
\tabletypesize{\scriptsize}
\tablenum{4}
\tablecaption{Calculated Orbital Ephemerides and Residuals for LTT~1445BC\label{tab:bc_eph}}
\tablehead{ 
\colhead{Date} & 
\colhead{$\theta_{\rm eph}$} & \colhead{$\rho_{\rm eph}$} & 
\colhead{$\Delta \theta$} & 
\colhead{$\Delta \rho$}\\
\colhead{} & 
\colhead{(deg)} & 
\colhead{(${\prime \prime}$)} &
\colhead{(deg)} & 
\colhead{(${\prime \prime}$)} 
} 
\startdata 
2003.4620  & 137.9 & 1.3452 &  0.2  & -0.0012\\
2008.7675  & 138.3 & 0.7166 & -0.6  &  0.0139 \\
2010.594   & 138.7 & 0.4320 & -0.3  & -0.0220\\
2012.0963  & 140.1 & 0.1786 &  1.7  &  0.0026\\
2012.0963  & 140.1 & 0.1786 & -0.1  & -0.0009\\
2012.7516  & 144.2 & 0.0648 &  2.3  &  0.0062\\
2012.7516  & 144.2 & 0.0648 & -2.8  &  0.0046\\
2014.7557  & 316.4 & 0.2821 & -0.3  &  0.0017\\
2014.7557  & 316.4 & 0.2821 &  0.2  &  0.0003\\
\enddata 
\end{deluxetable}

\section{Observations} \label{sec:data}

\subsection{TESS}

LTT~1445A and BC were observed by \tess ~in Sector four from UT 2018 October 19 to November 15, in spacecraft orbits 15 and 16. The observations were acquired with CCD 4 on Camera 2. We included this system in our \tess ~Guest Investigator program (PI Winters; G011231) target list to gather short-cadence (two-minute) data of the volume-complete sample of mid-to-late M dwarfs within 15 parsecs. LTT~1445A and BC were also included in the \tess ~Input Catalog (TIC) and Candidate Target List (CTL) \citep{Stassun(2018)} via the Cool Dwarf Sample \citep{Muirhead(2018)}.

The two-minute cadence data were reduced with the NASA Ames Science Processing Operations Center (SPOC) pipeline \citep{Jenkins(2015),Jenkins(2016)} that was repurposed from the $Kepler$ reduction pipeline \citep{Jenkins(2010)}. A planetary candidate with radius 1.4$\pm$0.4 \rearth ~was detected based on four transits to have a period of 5.4 days and a transit depth of 2498$\pm$168 ppm with a signal-to-noise ratio of 15.4. 

As noted in the data release notes\footnote{\url{ https://archive.stsci.edu/missions/tess/doc/tess_drn/tess_sector_04_drn05_v04.pdf}} for sector 4, this was the first sector to benefit from the improved Attitude Control System algorithm, which reduced the pointing jitter of the spacecraft by an order of magnitude over the pointing errors evident in data from sectors 1-3. Two anomalies were noted in sector 4. An incorrect guide star table was initially used; when the correct guide star table was uploaded, the spacecraft pointing shifted by 4\arcsec. All cameras show a maximum attitude residual of about 0.45 pixels that decreases to roughly 0.2 pixels once the guide star catalog was updated. In addition, communications between the spacecraft and instrument ceased for roughly 64 hours (between times 2458418.54 and 2458421.21), during which time no telemetry or data were collected.

\begin{figure*}
\centering
\includegraphics[scale=.70,angle=0]{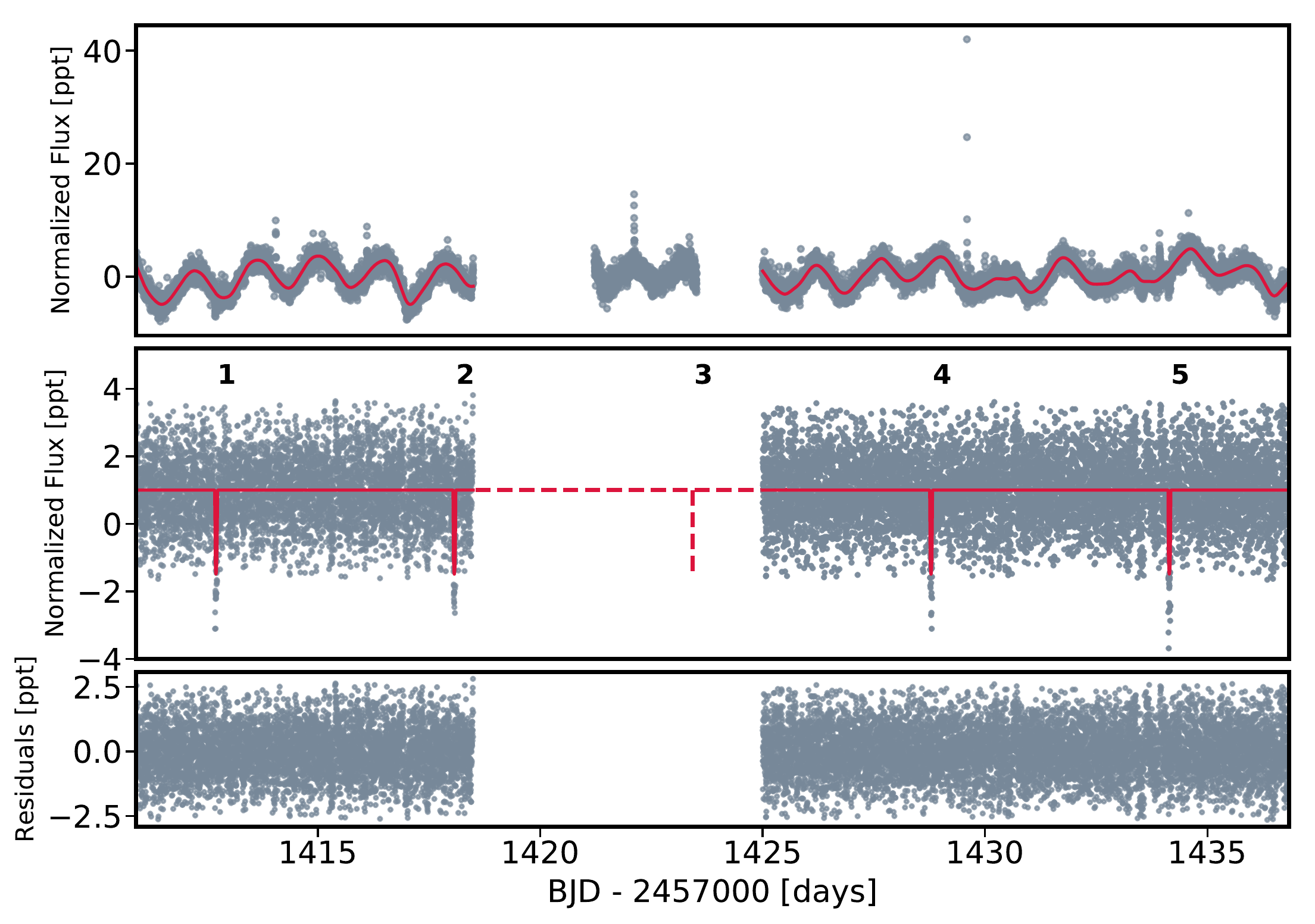}
\caption{The \tess ~PDCSAP light curve. The top panel shows evidence of flares and rotational modulation due to stellar spots, likely on either the B or C component. The solid (crimson) line indicates the fit to the modulation. The middle panel shows the residual data after the removal with Gaussian Processes (GP) of the stellar variability; the planetary transit model is overplotted (solid crimson line) and the number for each transit is indicated. The dotted line indicates the third transit, which was not included in our fit. The bottom panel illustrates the light curve with planetary transit and stellar variability removed. The description of the flare rejection and GP fitting are given in Section \S \ref{subsubsec:exoplanet}.  \label{fig:gp_lc}}
\end{figure*}

The \tess ~light curve, shown in Figure \ref{fig:gp_lc}, shows various types of stellar variability, such as flares and rotational modulation due to spots. We estimate a rotation period of 1.4~days from the \tess ~light curve (described in more detail in \S \ref{subsubsec:exoplanet}), which we suspect originates from either the B or C component, based on the activity indicator measurements for A and BC from our TRES spectra (described below). The duty cycle of the flares is fairly low, and those that occur are rather weak, with an increase in brightness on the order of 4\% (roughly 40 mmag); however, one large flare was detected, as seen in the top panel of Figure \ref{fig:gp_lc}, which was also reported by \citet{Howard(2019)}. While we do not know from which star the flares originate, previous work indicates that rapidly rotating stars flare more frequently. Thus, we suspect that the flares, too, come from either or both of the B or C components.

\subsection{Photometric Follow-Up with MEarth-South}
\label{subsec:mearth}

The large \tess ~pixel size of 21\arcsec ~means that the two nearly equal-luminosity sources, A and the blended BC, are included in the \tess ~aperture. Therefore it was necessary to determine from which star the transit signal originated. We obtained follow-up observations with MEarth-South for this purpose. One transit ingress of \planet ~was observed using four telescopes of the MEarth-South array at Cerro Tololo Inter-American Observatory (CTIO), Chile, on UT 2019 February 16.  Exposure times were 4 seconds, with a total of 1445 data points gathered over 3 hours at airmasses 1.2 to 3.0 starting immediately at twilight until the target set.  Due to the combination of the short exposure time and high airmass, particularly at transit ingress and thereafter, these data show very high levels of noise due to atmospheric scintillation.  Two of the telescopes used in this observation had shutters stuck in the open position, and as a result the images were smeared during readout, but this does not appear to affect the resulting differential photometry.

\begin{figure}
\includegraphics[scale=.40,angle=0]{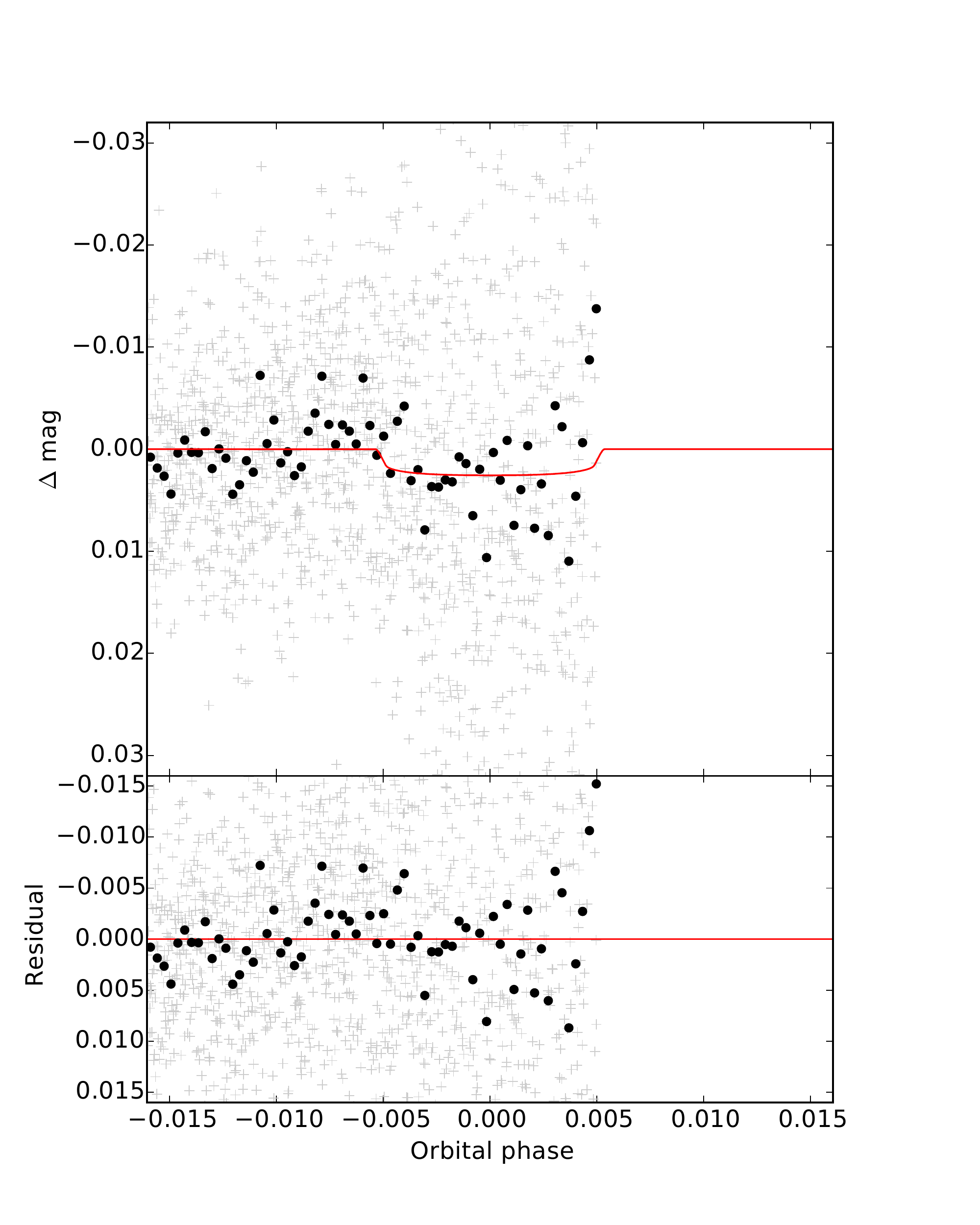}
\caption{The partial transit event of \planet ~(top panel) observed by MEarth-South, suggesting that the transit comes from \star. The data have been phase-folded using the final transit parameters given in Table \ref{tab:pparams}. The individual data points are indicated in gray; black points indicate the data binned by 2.5 minutes; the red line illustrates the fit to the transit event, computed by taking the mean of the MCMC posterior parameters from Table \ref{tab:pparams}. Error bars on the individual data points have been omitted for clarity, but are large due to the high airmass of the observation. Residuals to the fit are shown in the bottom panel.  \label{fig:mearth}}
\end{figure}

The FWHM of the stellar images ranged from approximately 2.0 to 3.5 pixels, with a plate scale of $0.84$ arcsec pixel$^{-1}$.  These observations resolve A from BC, but the wings of the point spread functions are still mildly overlapping and required specialized reduction procedures.  In order to mitigate the influence of aperture positioning errors, the global astrometric solutions for the images were used for aperture placement (e.g., as described by \citealt{Irwin(2007)}), rather than individually determining the location of each star from the individual images.  Undersized extraction apertures with radii of 4.2 pixels were used for the A and BC components, where the aperture size was chosen to prevent overlap of the apertures.  Due to the lack of useful comparison stars elsewhere on the images, BC was used as the comparison source to derive differential photometry of A.  The resulting light curve is shown in Figure \ref{fig:mearth}. It is compatible with the transits detected by \tess, although with low signal-to-noise, and suggests the transits originate from A. While we acknowledge that this detection is marginal, we provide further confirmation that the planet orbits the primary component in the system in \S \ref{sec:analysis}.

\subsection{Reconnaissance Data}

We used high-resolution data we had previously acquired as part of our ongoing, nearby M dwarf binary surveys \citep{Winters(2019a)} to confirm that the host star has no additional stellar or brown dwarf companions at separations less than 50 AU. We also investigated its rotational broadening and measured the equivalent width of H$\alpha$. In addition, we used roughly two decades of RECONS data from the CTIO / SMARTS 0.9m telescope to explore the system's long-term variability.

\subsubsection{High-Resolution Speckle Imaging: DSSI}

As part of our all-sky speckle survey of 1000 nearby M dwarfs for stellar companions (Winters, in prep), we observed \star ~on UT 2016 January 18 using DSSI on the Gemini-North 8.1m telescope. One thousand 40 millisecond exposures were taken in two filters (centered at 692 nm and 880 nm) simultaneously. This group of frames was followed by a similar set of exposures for a bright, unresolved calibration star at close proximity on the sky to the science star. The data reduction and analysis were conducted as described by \citet{Horch(2017)}. 



\begin{figure*}
\minipage{0.50\textwidth}
\includegraphics[scale=.45,angle=0]{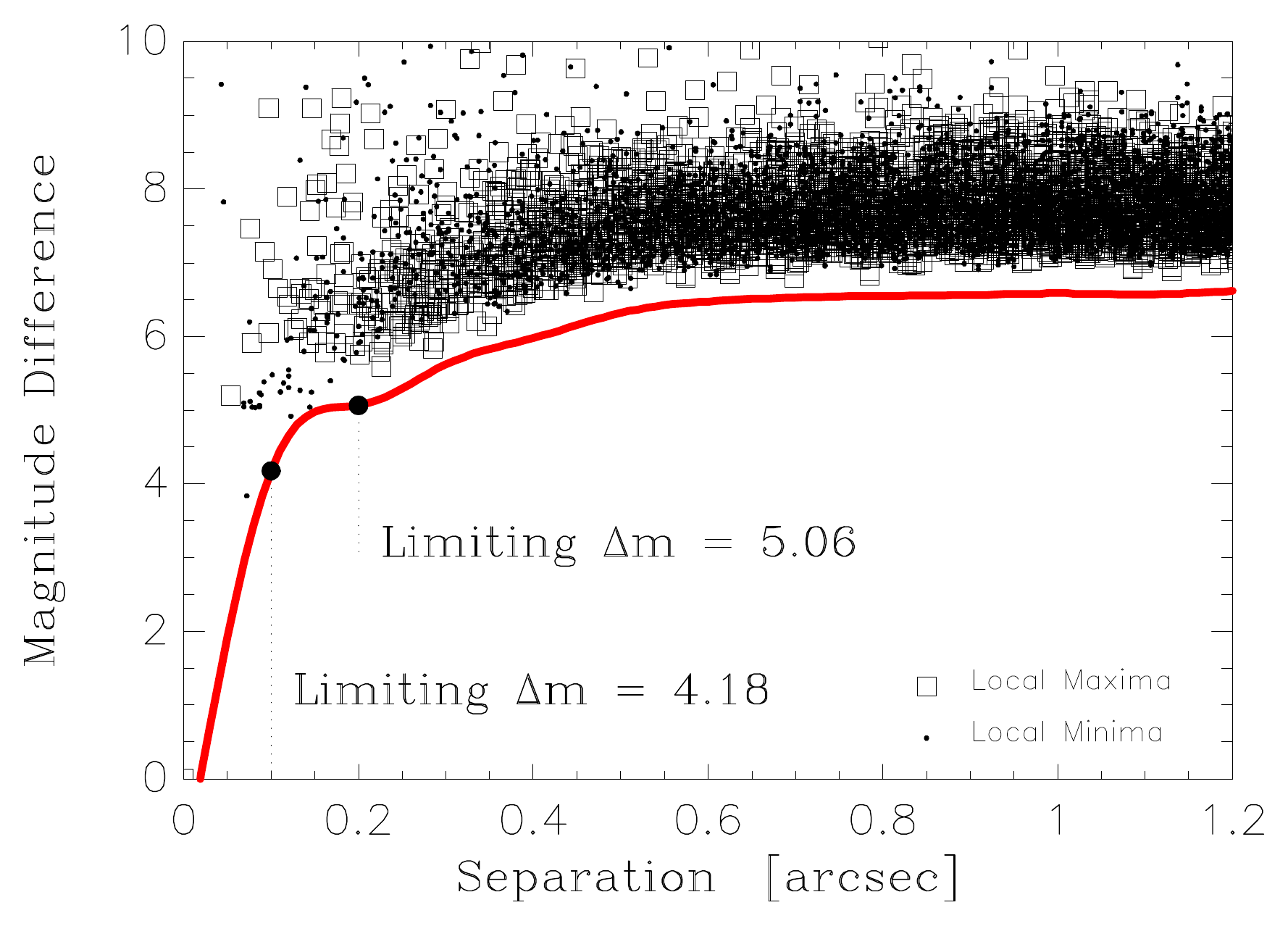}
\endminipage\hfill
\minipage{0.50\textwidth}
\includegraphics[scale=.45,angle=0]{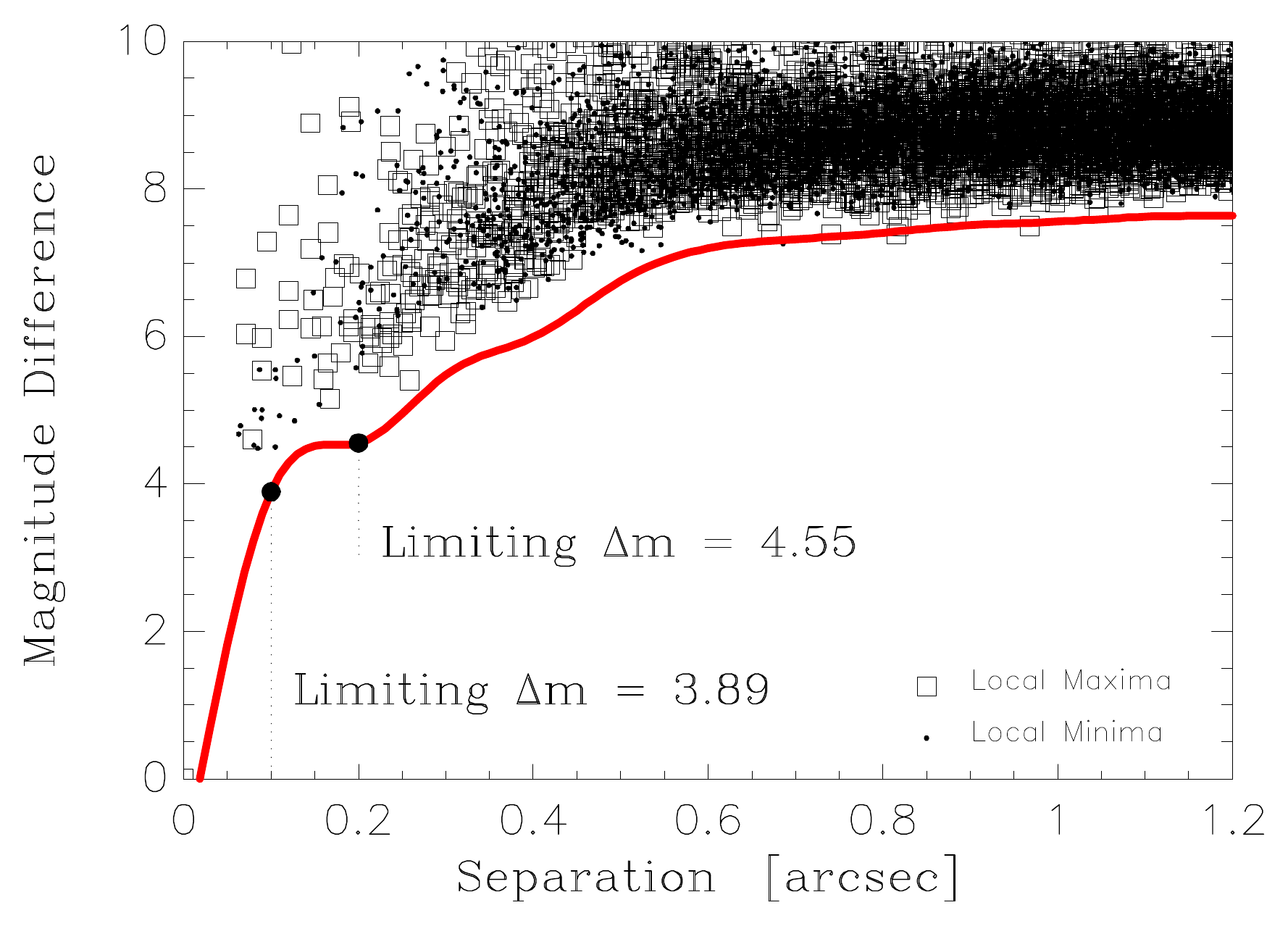}
\endminipage\hfill
\caption{Contrast curves of \star ~from DSSI on the Gemini-North 8.1m in the 692 nm (left panel) and 880 nm (right panel) filters. Open squares represent the positions of local maxima in the reconstructed image and points represent local minima (where the absolute value of the minimum is used). The red line represents the 5-$\sigma$ line as a function of separation. The lack of points below the red line illustrates that the primary component of LTT~1445 has no companions with $\Delta$mag less than 4.55 mag at separations 0\farcs2 -- 1\farcs2 (corresponding to projected linear separations of 1.4 -- 8.2 AU) from \star. \label{fig:speckle_ccs}}
\end{figure*}

We show the contrast curves for the 692 nm (left panel) and 880 nm (right panel) filters in Figure \ref{fig:speckle_ccs}. As illustrated in the left panel, no companions to \star ~were detected with  $\Delta$692nm less than 5.06 mag at separations 0\farcs2 - 1\farcs2 (corresponding to projected linear separations of 1.4 - 8.2 AU) from \star. Additionally, no companions were detected with $\Delta$880nm less than 7 mag at separations larger than 0\farcs6 (roughly 4.1 AU). An L2 V spectral type, what we consider to be the `end of stars' \citep{Dieterich(2014)}, has M$_I$ of roughly 16.0 mag. The M$_I$ for \star ~is 9.48 mag, placing constraints to M$_I$ of 16.48 mag on the presence of a companion at separations greater than 0\farcs6, beyond the end of the M dwarf sequence and into the brown dwarf regime. This is in agreement with results by \citet{Dieterich(2012)} who observed the system with \hst ~(NICMOS) and did not detect a stellar or brown dwarf companion to the primary star.


\subsubsection{High-Resolution Spectroscopy: TRES}

As part of our all-sky spectroscopic survey of mid-to-late M dwarfs within 15 pc, we acquired multiple high-resolution spectra of both the A and BC components with the Tillinghast Reflector Echelle Spectrograph (TRES) on the FLWO 1.5m Reflector. TRES is a high-throughput, cross-dispersed, fiber-fed, echelle spectrograph with a resolving power of roughly $R = 44\,000$ when using the medium fiber ($2\farcs3$ diameter) and a passband of 310-910~nm. The observations span UT 2017 February 03 to 2018 January 24, with four spectra of the primary and three of the BC pair acquired. We integrated for 120-600 and 120-180 seconds and achieved signal-to-noise ratios of 16-25 and 16-21 for A and BC, respectively, at 715 nm. We used the methods described in \citet{Winters(2018)} for our analysis. 

The equivalent width of the H$\alpha$ line and the magnitude of the rotational broadening allows differentiation between the two resolved stellar components. The primary star exhibits H$\alpha$ in absorption (we measure equivalent widths of 0.14$\pm$0.01\AA, 0.23$\pm$0.02\AA, 0.19$\pm$0.01\AA, and 0.25$\pm$0.01\AA), while the BC pair shows H$\alpha$ in emission (we measure equivalent widths of \mbox{-1.16}$\pm$0.03\AA, \mbox{-1.42}$\pm$0.03\AA, and \mbox{-1.52}$\pm$0.03\AA). We see negligible rotational broadening ($v$sin$i$) for the primary star, which allows us to place an upper limit of 3.4 km \pers ~(half the spectral resolution of TRES) on the rotational broadening of A; we measure a $v$sin$i$ of 4.4$\pm$3.4 km s$^{-1}$ for the blended secondary-tertiary pair. While we cannot rule out that the photometric modulation seen in the \tess ~light curve data comes from A, it is more likely to come from BC because we detect both H$\alpha$ in emission and rotational broadening in the spectra of those blended components. We use the use the relation $P_{rot} * v \sin i = 2 \pi R \sin i$, the 1.4-day rotation period, and the estimated radii for each component from Table \ref{tab:system_info}, to estimate rotational broadening of 9.7, 8.7, and 7.2 km \pers ~for the primary, secondary, and tertiary components, respectively, assuming each star is viewed edge-on.

We did not detect a second pair of lines in the spectra of the primary component that would indicate the presence of an additional stellar companion to A. Using the parameters from the preliminary astrometric orbit in Table \ref{tab:bc_orbit}, we calculate an expected radial velocity semi-amplitude of B due to C of 3.3 km \pers.



\subsubsection{CTIO / SMARTS 0.9m}

RECONS has been astrometrically and photometrically monitoring this system for over nineteen years at the CTIO / SMARTS 0.9m telescope. In fact, the first trigonometric parallaxes for A and BC were reported in \citet{Henry(2006)}. These data in the $V$ filter permit the investigation of the long-term variability of the system. Following the methods described in \citet{Hosey(2015)}, we note that, over the nineteen years of monitoring, the A component varies by 13.4 millimagnitudes (mmag) while the BC pair vary by 13.0 mmag.  While we do not consider either of these results indicative of a clearly variable source, for which we require an overall variability of 20 mmag, the levels for both are roughly twice that of the least variable M dwarfs at 7 mmag. We conclude that there are spots present on both A and the BC pair, but they do not change in coverage by an amount that alters the emergent flux in $V$ by more than 2\%. These results are in agreement with the moderate, short-term variability we detect in the \tess ~light curve data.

\subsection{HARPS}
\label{subsec:harps}

To constrain the mass of the transiting object, we acquired five new spectra of the primary star with the High Accuracy Radial Velocity Planet Searcher (HARPS) spectrograph \citep{Mayor(2003)} on the La Silla 3.6m telescope before the target was no longer observable. This totalled fourteen spectra when combined with the nine existing spectra in the ESO HARPS archive. HARPS is a fiber-fed echelle spectrograph with resolving power $R=$ 115,000 and a wavelength range of 378-691 nm. Spectra of LTT~1445A were integrated over 900 sec, except for the first two spectra acquired on 2003 December 15 and 2004 November 29, where the exposure times were 572 sec and 772 sec, respectively. The signal-to-noise ratio ranges between 37.8 and 72.5 at 650 nm, with an average of 55. For these spectra, this translates into radial velocity uncertainties \citep[computed following][and the procedure described below]{Bouchy(2001)} ranging between 0.9 m \pers ~and 1.7 m \pers, with an average uncertainty of 1.2 m \pers.

Radial velocities (RVs) were derived by a $\chi^2$-minimization, or so-called template matching. We briefly summarize here the implementation of the process, which is described in more detail in \citet{Astudillo-Defru(2015)}. We first used the RVs from the HARPS pipeline \citep{Lovis(2007)}, in combination with the barycentric corrections, to shift all spectra of \star ~to the solar system barycentric reference frame. We constructed an initial stellar template by computing the median of the shifted and stacked spectra. A telluric template was constructed by Doppler-shifting the observed spectra of \star ~to the laboratory rest frame, which aligns the telluric absorption features, and computing the median. An improved stellar template was then constructed with the known telluric lines removed; that is, the template is a true spectrum of the star itself with improved signal-to-noise ratio. Due to the low number of spectra and to prevent auto-correlation between the stellar template and the spectrum being analyzed, contrary to \citet{Astudillo-Defru(2015)}, we computed the stellar template for each epoch but discarded the spectrum under analysis. New RVs were then derived by minimizing the $\chi^2$ of the residuals between the observed spectra and the stellar template. We list the RVs in Table \ref{tab:harps_rvs}. 

The RVs of A exhibit a long-term drift due to the presence of the BC stellar pair, as shown in the bottom panel of Figure \ref{fig:harps_rvs}. This is in agreement with the astrometric perturbation of A due to BC mentioned above in \S \ref{sec:host_system}. These precise radial-velocity data also allow us to rule out the presence of any other nearby stellar or brown dwarf object orbiting the primary star. We note that observations will continue when the system is again observable to provide a robust mass for the planet.

\begin{deluxetable}{lcc}
\tabletypesize{\small}
\tablecaption{HARPS Radial Velocities for LTT~1445A  \label{tab:harps_rvs}}
\tablecolumns{3}
\tablenum{5}

\tablehead{
\colhead{BJD\tablenotemark{a}} & 
\colhead{$v_{\rm rad}$\tablenotemark{b}} &
\colhead{$\sigma$} \\ 
\colhead{(days)} & 
\colhead{(${\rm km\ s^{-1}}$)} &
\colhead{(${\rm km\ s^{-1}}$)} 
}
\startdata
2452988.689447  & -5.38629  &    0.00174\\
2453338.681357  & -5.39562  &    0.00116\\
2454078.640473  & -5.40353  &    0.00105\\
2454080.657212  & -5.40513  &    0.00108\\
2454292.913009  & -5.41529  &    0.00091\\
2454316.868711  & -5.41280  &    0.00112\\
2455042.915163  & -5.42530  &    0.00088\\
2455997.500569  & -5.43569  &    0.00137\\
2456237.681670  & -5.43910  &    0.00130\\
2458546.500296  & -5.45774  &    0.00131\\
2458547.501338  & -5.46353  &    0.00116\\
2458548.500633  & -5.46085  &    0.00136\\
2458555.506869  & -5.45518  &    0.00129\\
2458556.505906  & -5.45354  &    0.00116\\
\enddata
\tablenotetext{a}{Barycentric Julian Date of mid-exposure, in the TDB time-system.}
\tablenotetext{b}{Barycentric radial velocity.}

\end{deluxetable}


\section{Analysis}
\label{sec:analysis}

\subsection{Additional Confirmation of \star ~as the Host Star} \label{subsec:centroid}

We explored a number of other methods to determine from which star the transit originates. 


Because our system has high proper motion, we were able to investigate whether the host star had moved on top of a background star which could be the source of the transit signal. We compared the position of the primary star in digitally-scanned SuperCOSMOS \citep{Hambly(2001)} archival POSS-1 images (taken 1953.93) with the star's position at the time of the \tess ~observations (we chose November 1 -- 2018.83 -- the effective mid-point of sector 4). The system had moved 29\farcs6 since the POSS-1 image was taken. No background star was seen at the current position of \star ~in the POSS-1 image, which has a magnitude limit of roughly 19.5 mag at 620-670 nm.

The data validation (DV) report from the SPOC pipeline provides a number of tests to aid in analyzing planet candidates. Two of the tests in the DV report use a difference image to analyze the centroid shift that occurs during the transit event to determine from which star the transit signal originates. The difference image provides the flux value for each pixel by taking the difference between the mean out-of-transit flux value and the mean in-transit value. Therefore, the star that is the planet host will have residual flux in the difference image. A centroid is then determined for the difference image. One analysis compares the difference image centroid to the expected position based on the TIC coordinates, while a second analysis compares the centroid of the difference image to the out-of-transit centroid. For \system, the difference image centroid corresponds to LTT~1445A in both instances; however, the magnitudes of the two centroid shifts in the DV report disagree by more than a factor of three (an offset of 2\farcs969$\pm$0\farcs440 for the TIC coordinate centroid, compared to an offset of 10\farcs130$\pm$0\farcs414 for the out-of-transit centroid), so we investigated further. In contrast to the way this same analysis was conducted for \kepler ~objects of interest, the TIC coordinates are corrected for proper motion. Therefore, in crowded fields, the \tess ~centroid offset measurements with respect to the TIC coordinates are generally more reliable than the out-of-transit centroid. But as noted by \citet{Twicken(2010),Stumpe(2014),Twicken(2018)}, the out-of-transit centroid is subject to crowding and can lock on to a star that is not the target. That is, in fact, the case for \system: the out-of-transit centroid position corresponds to that of the BC pair instead of the A component, resulting in an overestimated centroid offset in the DV report. Adjusting the 10\farcs130 out-of-transit centroid offset by 7\farcs10, the most recently measured separation of A and BC, gives an offset of 3\farcs03, in agreement with the 2\farcs969 offset from the TIC coordinate centroid. 

In addition to the centroid shift tests, the candidate transit signature passes all the other diagnostic tests intended to flag false positives. These tests include the odd/even transit depth test, the weak secondary test, the ghost diagnostic test (which often flags background eclipsing binaries or scattered light features), and the statistical bootstrap test (false alarm probability $<$3e-16).

The field is sparse. The TFOP-TESS entry for this system reports ten contaminating sources, but only three additional sources are listed in the DV report and are shown to lie within the target mask. One is the nearby, physically bound pair which we have discussed above (LTT~1445BC, TIC 98796342). The other two sources are faint, with reported $T$ magnitudes of 15.032 and 15.991 at angular separations of 104\farcs51 and 120\farcs16 for TIC IDs 98796341 and 98796339, respectively. Our MEarth observations produced light curves for these two stars which confirm that they are not the source of the transit. 

Based on the above analyses, and in combination with our marginal ground-based MEarth detection, we are confident that the planet candidate is transiting the primary star in the system. 


\subsection{Host Star Parameters}
\label{subsec:star_params} 

We use the methods appropriate for M dwarfs previously used by our group \citep{Berta-Thompson(2015),Dittmann(2017a),Ment(2019)} to determine the stellar parameters of the planet candidate host, which we then used as priors for the light curve modeling described below. We estimate the mass of the host star using the mass-luminosity relation in the $K-$band by \citet{Benedict(2016)} to be 0.258$\pm$0.014 \mdot. The relation in $M_K$ has been found to be less sensitive to metallicity than the $M_V$ relation \citep{Henry(1993),Delfosse(2000),Benedict(2016)}. For comparison, the estimated mass from the $M_V$-band relation is 0.251$\pm$0.023 \mdot. We then use single-star mass-radius relations \citep{Boyajian(2012)} to find a stellar radius of 0.268$\pm$0.027 R$_{\odot}$. We calculate the bolometric correction in $K$ using the prescription in \citet[erratum]{Mann(2015)} to be 2.70$\pm$0.04 mag, resulting in a bolometric luminosity for \star ~of \mbox{0.0079$\pm$0.0003 L$_{\odot}$}. We calculate the correction in $V$ from \citet{Pecaut(2013)} to be -2.06$\pm$0.04 mag\footnote{We assume the uncertainty on the bolometric correction in $V$ is that of the ($V-K$) color.}, resulting in a bolometric luminosity of \mbox{0.0082$\pm$0.0004 L$_{\odot}$}. We adopt the mean of the two bolometric luminosities. From the Stefan-Boltzmann Law, we find an effective temperature $T_{\rm eff}$ of 3337$\pm$150 K. As a comparison, we also used the relations in \citet{Mann(2015)} to determine an effective temperature of 3332$\pm$77 K for \star, in agreement with the $T_{\rm eff}$ derived from the Stefan-Boltzmann Law. We adopt the [Fe/H] of -0.34$\pm$0.08 from \citet{Neves(2014)}, which is measured from HARPS data.   


\subsection{Light Curve Modeling}

Because of the complex nature of the light curve, we used a combination of {\sc exoplanet} \citep{Foreman-Mackey(2017)} and {\sc EXOFASTv2} \citep{Eastman(2017)} for our light curve modeling. The purpose of using {\sc exoplanet} was to fit and remove the photometric modulation in the \tess ~light curve using Gaussian Processes (GP) while preserving the planetary transit signal. We then used {\sc EXOFASTv2}, which does not currently have GP capability, to simultaneously fit the de-trended transit data from {\sc exoplanet} and the HARPS RV data. {\sc exoplanet} also has the capability to fit RV data, but we did not get convergence when including the HARPS data in the model. We elected not to include the third transit in our analysis because the light curve baseline showed a strong slope at egress; we note that this transit was also omitted from the results in the SPOC DV report.

If there are other objects in the \tess ~aperture with TIC identifiers, the SPOC pipeline calculates a dilution (i.e., contamination) factor and performs a correction to the final light curve, as noted in the $Kepler$ manual \citep{Thompson(2016)}. The correction for this system, contained in the keyword `CROWDSAP' in the light curve file, is 0.485. However, the \tess ~magnitudes of 8.88 and 8.80 that we calculate for each component are slightly different from the magnitudes in the TIC and CTL (for TIC 98796344(A), $T =$ 8.64 mag; for TIC 98796342(BC), $T =$ 8.55 mag) because of our M-dwarf-specific relation; thus, our calculated  dilution factor will be slightly different, as well. From our \tess ~magnitudes, we calculated the flux for each component, from which we determined the dilution ($f_A$ / ($f_A$ $+$ $f_B$ $+$ $f_C$)) to be 0.480$\pm$0.013. 


\subsubsection{{\sc exoplanet}}
\label{subsubsec:exoplanet}

For the first part of the light curve analysis, we used the python package {\sc exoplanet} \citep{exoplanet:exoplanet}. {\sc Exoplanet} employs probabilistic methods to model exoplanet transit and radial velocity data sets. It has the additional capability to incorporate Gaussian Processes (GP) with {\sc celerite} \citep{Foreman-Mackey(2017)} and limb-darkened light curves with {\sc starry} \citep{Luger(2018)}.  We used the SPOC-generated Pre-Search Data Conditioning Simple Aperture Photometry (PDCSAP) light curve \citep{Smith(2012),Stumpe(2014)}, corrected with our calculated dilution factor. Before fitting, we removed positive outliers (flares) deviating by more than 3.0-$\sigma$ above the median absolute deviation of the PDCSAP light curve. To model the planetary transit, we used a limb-darkened transit model and a Keplerian orbit. The stellar variability, as well as any other possible systematics, are modeled with a GP.


We parameterized the model by the radius of the star in solar units $R_{*}$, mass of the star in solar units $M_{*}$, time of transit $T_0$ in days, orbital period $P$ in days, transit impact parameter $b$, eccentricity $e$, and argument of periastron $\omega$. This is used as the input for the light curve modeler, {\sc starry}, which computes a limb-darkened light curve, with parameters for quadratic limb-darkening coefficients $u_1$ and $u_2$. In addition to the limb-darkening coefficients, we parameterized the {\sc starry} light curve by the model mentioned above, the radius of the planet $R_{p}$, the times for which the light curve is to be evaluated, the exposure time of each observation, which in our case is 120 seconds, and the mean of the stellar flux, $F_{*_{\mu}}$ in parts per thousand (ppt). We performed a Box-Least-Squares (BLS) periodogram analysis on the PDCSAP light curve in order to estimate $P$, $T_0$, and the transit depth $\delta$. These estimates were used to inform the priors for $P$ and $T_0$. We used the transit depth as a constraint on the Gaussian prior placed on the radius of the planet. The priors are summarized in Table \ref{tab:PLpriors}.


\begin{deluxetable}{lccc}
\tabletypesize{\small}
\tablecaption{\sc{exoplanet} Planetary Orbit Parameters  \label{tab:PLpriors}}
\tablecolumns{4}
\tablenum{6}

\tablehead{
\colhead{Parameter} & 
\colhead{Prior\tablenotemark{a,b}} & 
\colhead{Value} &
\colhead{Bound} 
}
\startdata
$R_{*}$ ($R_{\odot}$)   &  Gaussian   & $\mu =$ 0.268  & $\sigma$ = 0.013  \\
$M_{*}$ ($M_{\odot}$)   &  Gaussian   & $\mu =$ 0.258  & $\sigma$ = 0.014  \\
$F_{*_\mu}$ (ppt)       &  Gaussian   & $\mu =$ 0.0    & $\sigma$ = 10.0  \\
$u_1$                   & Flat        & 0.0            & (0.0 - 1.0) \\ 
$u_2$                   & Flat        & 0.0            & (0.0 - 1.0) \\
$T_0$ (days)            &	Gaussian  & $\mu =$ -0.448 & $\sigma$ = 0.5 \\
$\rm{log}~P$ (days)     &	Gaussian  & $\mu =$ ~\rm{log}$~5.358$ 	& $\sigma$ = 0.5 \\
$b$                     &	Uniform   &  0.5	       & (0.0 - 1.0) \\
$e$                     &	Beta     & $\alpha =$ 0.876, $\beta =$  3.03 & (0.0 - 1.0) \\
$\omega$                & Uniform	  & 0.0            & ($-\pi$ - $+\pi$)\\
$\rm{log}~R_p$          & Gaussian    & $\mu =$ 0.5($\rm{log}~\delta + \rm{log}~R_{*}$) & $\sigma$ = 1.0\\
\enddata
\tablenotetext{a}{Where $\mu$ and $\sigma$ parameterize the Gaussian distribution.}
\tablenotetext{b}{Where $\alpha$, and $\beta$ parameterize the Beta distribution.}
\end{deluxetable}


The GP kernel is the sum of two simple harmonic oscillators shown in Equations \ref{eq:sho1} and \ref{eq:sho2}:


\begin{equation}
\label{eq:sho1}
    SHO_1(\omega_{GP}) = \sqrt{\frac{2}{\pi}} \frac{S_1\,\omega_1^4}
        {(\omega_{GP}^2-{\omega_1}^2)^2 + {\omega_1}^2\,\omega_{GP}^2/Q_1^2} 
\end{equation}

and

\begin{equation}
\label{eq:sho2}
       SHO_2 (\omega_{GP}) = \sqrt{\frac{2}{\pi}} \frac{S_2\,\omega_2^4}
        {(\omega_{GP}^2-{\omega_2}^2)^2 + {\omega_2}^2\,\omega_{GP}^2/Q_2^2}
\end{equation}

where, 

\begin{equation}
    Q_1 = 0.5 + \tau_3 + \tau_4
\end{equation}

\begin{equation}
    \omega_1 = \frac{4\pi Q_1}{\tau_2 \sqrt{4Q_1^2 - 1}}
\end{equation}

\begin{equation}
    S_1 = \frac{\tau_1}{\omega_1 Q_1}
\end{equation}

\begin{equation}
    Q_2 = 0.5 + \tau_3
\end{equation}

\begin{equation}
    \omega_2 = \frac{8\pi Q_2}{\tau_2 \sqrt{4Q_2^2 - 1}}
\end{equation}

\begin{equation}
    S_2 = \frac{\tau_5~\tau_1}{\omega_2 Q_2}
\end{equation}

This is an appropriate kernel for data that are quasi-periodic in nature, such as the observed rotational modulation in the light curve of LTT~1445ABC. The hyper-parameters for this GP are the amplitude of variability $\tau_1$, the primary period of the variability $\tau_2$, the quality factor $\tau_3$, the difference between the quality factors of the first and second modes of the two oscillators $\tau_4$, the fractional amplitude of the secondary mode to the primary mode $\tau_5$, and a jitter term added to account for excess white noise $\tau_6$. We placed a uniform prior on $\tau_5$ and Gaussian priors on $\tau_1$, $\tau_2$, $\tau_3$, $\tau_4$, and $\tau_6$. The mean ($\mu$) and standard deviation ($\sigma$) value we set  for the Gaussian prior on $\tau_1$, $\tau_2$, $\tau_3$, $\tau_4$, and $\tau_6$ are shown in Table \ref{tab:GPpriors}. For the Gaussian prior on $\tau_2$, the primary period of variability, we estimated the mean, equal to 1.4 days, using a Lomb-Scargle periodogram of the light curve with the transits masked out.

\begin{deluxetable}{lccc}
\tabletypesize{\small}
\tablecaption{Gaussian Process Hyper-parameters  \label{tab:GPpriors}}
\tablecolumns{4}
\tablenum{7}

\tablehead{
\colhead{Hyper-parameters} & 
\colhead{Prior} & 
\colhead{Value} &
\colhead{Bound} 
}
\startdata
$\rm{log}~\tau_1$ (ppt)  &  Gaussian &  $\mu =${\rm{log}~var(Flux)} 	& $\sigma$ =  5.0  \\
$\rm{log}~\tau_2$ (days) &	Gaussian & $\mu = ${\rm{log}(1.4)} 	        & $\sigma$ =  1.0  \\
$\rm{log}~\tau_3$        &	Gaussian & $\mu = ${~\rm{log}(-5.0)}         & $\sigma$ = 1.0 \\
$\rm{log}~\tau_4$        &	Gaussian & $\mu = ${~\rm{log}(-5.0)}	        & $\sigma$ =  2.0\\
$\tau_5$                 &	Uniform  &  0.0	                            & (0.0 - 1.0) \\
$\rm{log}~\tau_6$ (ppt)  &	Gaussian &  $\mu = ${~\rm{log}~var(Flux)} 	& $\sigma$ =  10.0 \\
\enddata
\end{deluxetable}

We implemented the GP-transit model using {\sc PyMC3} \citep{exoplanet:pymc3}. Before sampling the model, we compute an initial guess of the parameters using a built in optimizer from {\sc exoplanet}. The optimization finds the maximum a posteriori solution, which is used as starting values for the sampler. After an initial burn-in of 14,000 steps, 12,000 steps were drawn from the posterior. {\sc PyMC3} provides useful convergence diagnostics such as the Gelman-Rubin statistic and the number of effective samples. For each parameter in our joint GP-transit model, the Gelman-Rubin statistic was within 0.001 of 1.000 and the number of effective samples was above 4000. We show the results of our fit in Figure \ref{fig:gp_lc}.
 



\subsubsection{{\sc EXOFASTv2}}

We used the output light curve data from {\sc exoplanet} with the stellar variability removed as input for the software package {\sc exofastv2} \citep{Eastman(2013),Eastman(2017)}. {\sc exofastv2} is a suite of {\sc IDL} routines that simultaneously fits exoplanetary transit and radial velocity data using a differential Markov Chain Monte Carlo (MCMC) code. 

Because we derived the stellar parameters as described in \S \ref{subsec:star_params}, we did not include a spectral energy distribution in the fit, and we disabled the default MIST stellar evolutionary models that use isochrones to constrain the stellar parameters. We placed Gaussian priors on the mass, radius, effective temperature, and metallicity that were equal to the uncertainties noted in \S \ref{subsec:star_params}. The quadratic limb darkening coefficients were constrained by the \tess ~data and penalized for straying from the values predicted by the \citet{Claret(2017)} limb darkening tables at a given log{\it g}, $T_{\rm eff}$, and [Fe/H], as is standard within EXOFASTv2. While the atmospheric models used to derive the limb darkening tables are questionable for low-mass stars such as \star, the impact is likely to be negligible due to the low precision of the \tess ~lightcurve. 


As noted in \S \ref{subsec:harps}, the fifteen years of HARPS RVs exhibit a drift due to the presence of the stellar BC pair, so we included terms for the slope and quadratic curve of the RVs. Additionally, we modified the default that searches in logarithmic RV semi-amplitude space to a linear option because of the few RV measurements available. We allowed eccentricity to be a free parameter, but excluded values where eccentricity was greater than $1 - 3 R_*/a$. Tides would theoretically be expected to exclude such high eccentricities because the tidal circularization timescale is very short \citep{Adams(2006)}. In addition, the excluded eccentricities are at negative RV semi-amplitudes, which omits non-physical masses for the planet. In order to allow for the propagation of the uncertainty in the dilution which was corrected in the {\sc exoplanet} fit, we included a prior on the correction to the dilution that was a Gaussian centered on zero with $\sigma = 0.013$. We required the number of independent draws to be greater than 1000 and and determined that, with a Gelman-Rubin statistic of 1.0107 in the worst case, the chains were well-mixed. 

\begin{figure}
\includegraphics[scale=.80,angle=0]{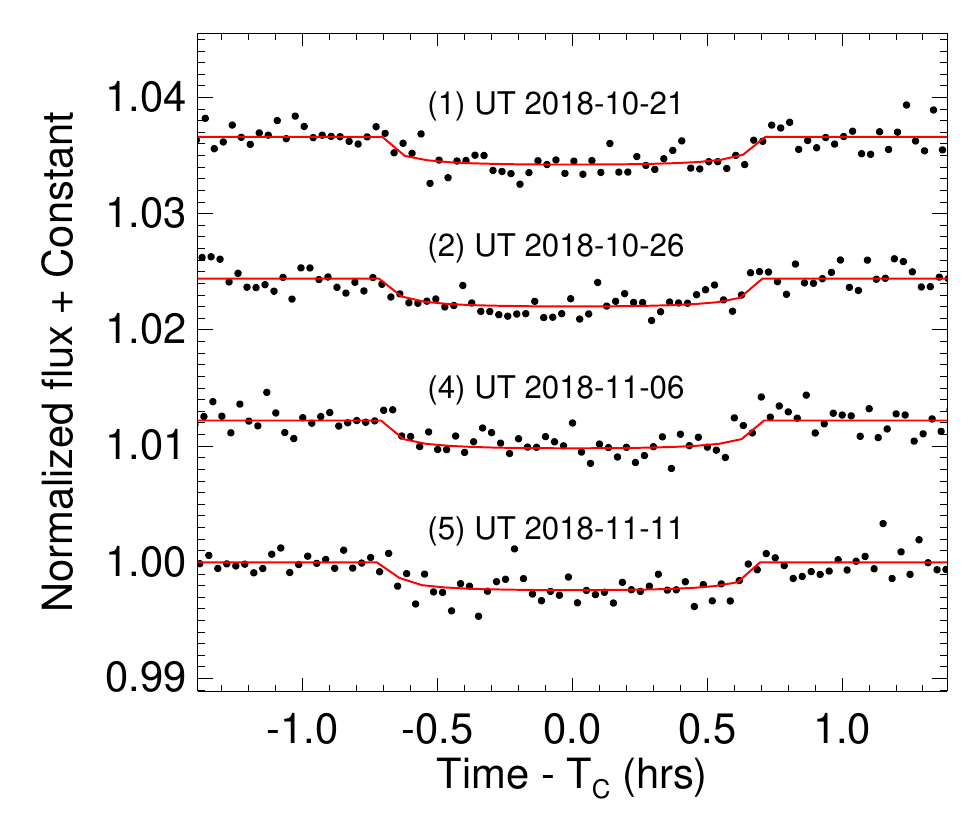}
\includegraphics[scale=.80,angle=0]{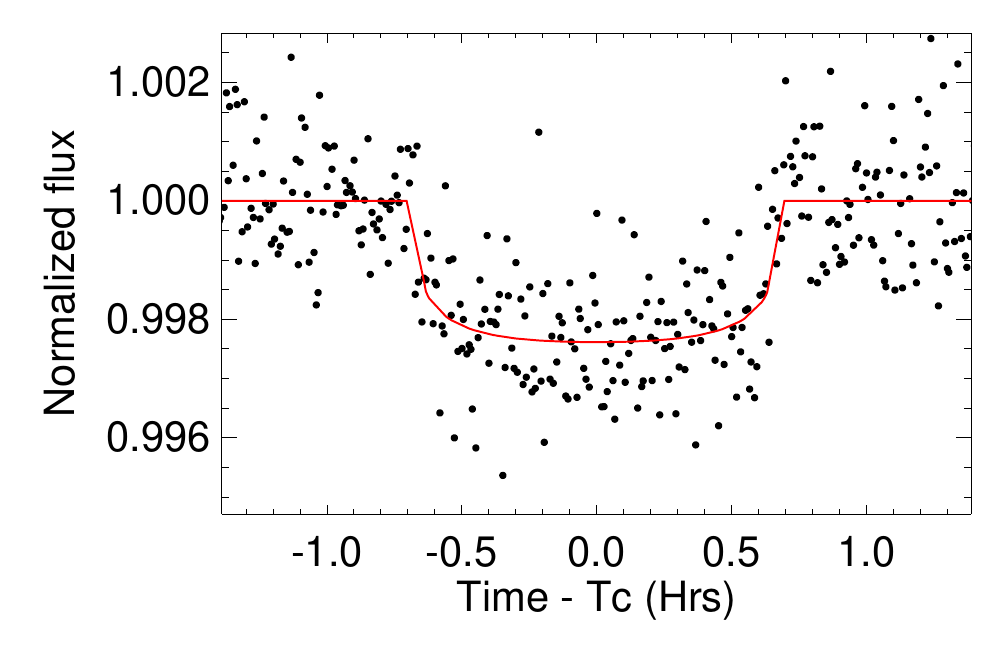}
\caption{The transits of \planet ~from \tess ~data, fit simultaneously with the RV data using {\sc exofastv2}. The individual transits are shown in the top panel, where the numbers in parentheses correspond to the transit numbers in Figure \ref{fig:gp_lc}; the phase-folded transit is shown in the bottom panel. \label{fig:transits_tess}}
\end{figure}

\begin{figure}
\includegraphics[scale=.80,angle=0]{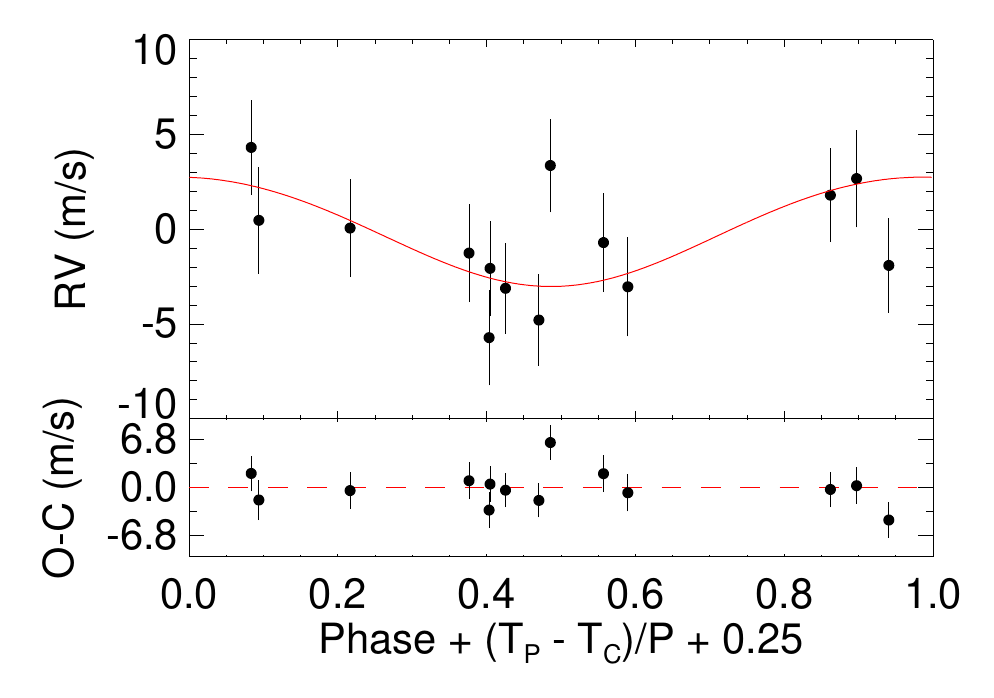}
\includegraphics[scale=.80,angle=0]{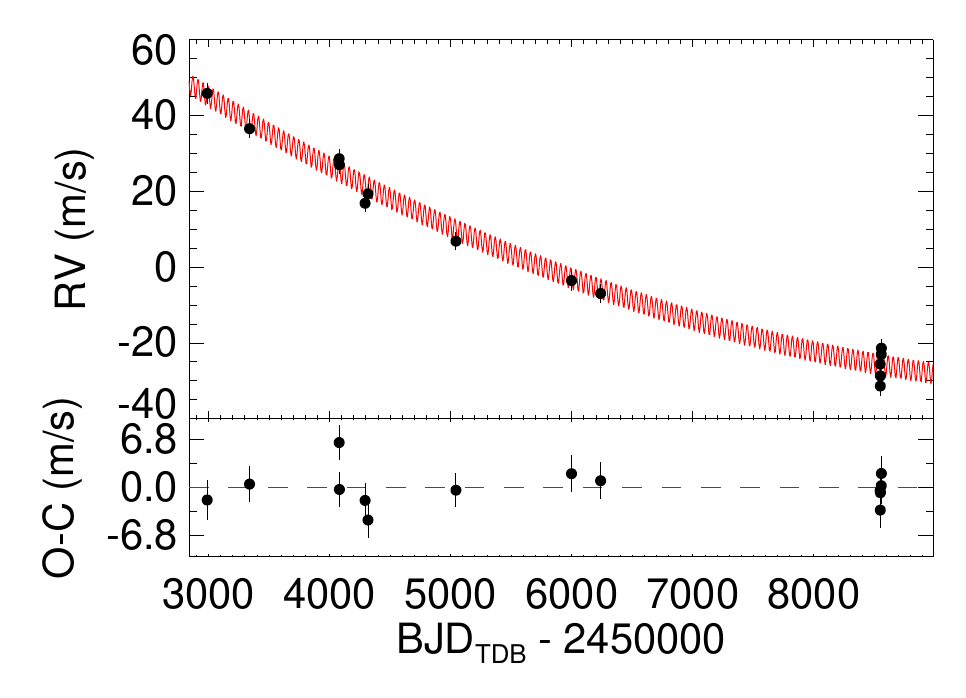}
\caption{The preliminary spectroscopic orbit of \planet ~from HARPS data, fit simultaneously with the \tess ~transit data using {\sc exofastv2}. The orbit and its residuals are shown in the top panel, while the velocity drift of the primary star due to the stellar BC components is evident in the bottom panel, along with the residuals. \label{fig:harps_rvs}}
\end{figure}

We find a period of \pplanet ~days, radius of \rplanet ~\rearth, mass of \mplanet ~\mearth, and equilibrium temperature of \teqplanet ~K for \planet. The equilibrium temperature assumes an albedo of zero with perfect redistribution. We show the individual transits, along with the phase-folded transit in Figure \ref{fig:transits_tess}. In Figure \ref{fig:harps_rvs}, we show the best-fit model to the RV data, which we acknowledge is marginal. The uncertainties illustrated are the RV uncertainties from Table \ref{tab:harps_rvs} added in quadrature with the fitted RV jitter. The uncertainty on the planet's mass is largely due to the sparse RV coverage, so we place a 3-$\sigma$ upper limit on the RV semi-amplitude K, planet mass, and planet density. The 3-$\sigma$ upper limit is where 99.73\% of all links of all chains, after discarding the burn-in, are smaller than the quoted value.  Listed in Table \ref{tab:pparams} are the 68\% confidence values from the {\sc exofastv2} fit. In addition we list the 3-$\sigma$ upper limits for K and its derived parameters.

To confirm that the two light curve fitting packages produce the same result, we also ran {\sc exofastv2} on the GP-corrected transit data without including the radial velocity data. The transit-only results from {\sc exofastv2} and {\sc exoplanet} are consistent with each other, within the errors. We also compared results from a fit to the detrended, whitened (i.e., the data validation time series) \tess ~light curve to our {\sc exoplanet} $+$ {\sc exofastv2} fit results and found good agreement.



\section{Discussion} 
\label{sec:disc}

To summarize, we have presented the discovery of \planet, which resides in a host system composed of three mid-to-late M dwarfs at 6.9 pc. The planet has a radius \rplanet ~\rearth, an orbital period of \pplanet ~days, and an equilibrium temperature of \teqplanet ~K; we place a 3-$\sigma$ upper mass limit of \muplanet ~\mearth ~on the planet. We have also presented a detailed view of the host system, which includes a preliminary orbit for the bound BC stellar pair that shows it to be in an eccentric and edge-on configuration. 

The planet is an S-type (satellite) planet, meaning that it orbits one component of a stellar multiple system \citep{Dvorak(1982)}. Most of the literature on planets in multiple star systems considers planets in binary systems with solar-type primary stars \citep{Wang(2014a),Wang(2014b),Winn(2015),Kraus(2016),Eggenberger(2010),Matson(2018)}. Partly, this is because attention has only recently shifted to M dwarfs as planet hosts. Because stellar multiplicity is known to be a decreasing function of primary mass \citep{Duchene(2013)}, M dwarfs have a smaller stellar multiplicity rate than more massive stars:  \citet{Winters(2019a)} estimate it to be 26.8$\pm$1.4\%, in agreement with recent results of 26$\pm$3\% and 28.6$^{+2.7}_{-3.1}$\% from \citet{Duchene(2013),Ward-Duong(2015)}, respectively. As a result, there are fewer systems to discuss, even if every M dwarf system has a planet. Theoretical work has shown that planets in binary systems are in stable configurations if their separations from their host stars are less than one-third the distance to a gravitationally-bound companion \citep{Holman(1999)}. The 0.038 AU separation of \planet ~is significantly less than one-third the 21 AU separation between A and BC (the smallest separation measured between A and B by Rossiter in 1943). Therefore, the planet is likely in a dynamically stable orbit, even given the presence of two other stars in the system.

M dwarfs are thought to be more likely than more massive stars to host multiple planets in compact orbits of less than roughly ten days \citep[ respectively, for early and mid-to-late M dwarfs]{Ballard(2019),Muirhead(2015)}. LTT~1445A may very well host multiple planets, but with only one sector of data from \tess, we have not yet seen indications of other planets in the system. 

The alignment of the three stellar components and the edge-on orbit of the BC pair and planet is suggestive of the co-planarity of the system. Including LTT~1445ABC, there are six `pure' M dwarf triples known within 10~pc.\footnote{We do not include GJ~799/803 or GJ~569ABC. We consider the primary of the GJ~799/803 system to be a late K-type star, based on the M dwarf color cuts described by \citet{Winters(2019a)}, and we consider one of the companions to GJ~569A to be a brown dwarf, again consistent with \citet{Winters(2019a)}. Further, we do not include M dwarf triples that are subsets of higher-order multiples.} Among these are the very tight systems LTT~12352 and GJ~866 that have separations much smaller than LTT~1445. The remaining three systems --- GJ~2005, GJ~1230, and GJ~1245 --- all have archival \hst ~NICMOS images available. Remarkably, none appear to exhibit the possible co-planarity we observe in \system. The presence of a transiting planet in this system raises the possibility that the entire system is co-planar, which has intriguing implications for planet formation scenarios. 


Young stars form in often densely clustered environments with separations between the stars on the order of hundreds of astronomical units. Dynamical evolution of the cluster could result in capture, leading to binaries with wide separations; however, these stellar components would likely not be co-planar. A more likely formation mechanism is the quasi-static fragmentation of the circumbinary disk as the protostellar core is collapsing, where conservation of angular momentum leads to the formation of an accretion disk \citep{Stahler(2005)}.  Such a system would possess both circumbinary and circumstellar disks. At apastron, a bound stellar pair would disrupt the circumbinary disk, which would tend towards increased eccentricity in the orbit of the stellar pair. Truncation of any circumstellar disks, and the cessation of accretion onto the primary's disk could truncate both the stellar mass and the masses of any planets around the primary. Given the possible co-planarity of the LTT~1445 system, the low masses of the stellar components, the eccentricity of the orbit of the BC stellar pair, and the presence of the terrestrial planet, it is likely that the system formed from the early fragmentation of an individual protostellar core, and not from later dynamical evolution within the young stellar cluster in which the stars formed.

Previous work leads us to expect that \planet ~is rocky in composition. Planets with radii less than 1.4 R$_{\oplus}$ are typically terrestrial in nature \citep{Rogers(2015),Dressing(2015a)}. An Earth-like density for \planet ~in a circular orbit implies a planetary mass of 2.5 \mearth ~and an RV semi-amplitude of 2.3 m \pers. We expect the mass of the planet to be below \muplanet ~\mearth, based on our HARPS data. Additional precise RV observations in the near future will provide a robust mass for the planet; this will allow refinement of the planet's surface gravity, which will serve as an input for atmospheric models. 

The planet is not in the habitable zone (HZ) of its star. The conservative inner and outer HZ boundaries for a 1~M$_{\oplus}$ planet around a star with $T_{\rm eff}$ of 3335~K are 0.093 and 0.182 AU, respectively, as calculated from the relations in \citet{Kopparapu(2013),Kopparapu(2014)}. The planet receives 5.6 times the Sun's irradiance from its host star, as its orbital distance brings it closer to \star ~than the annulus of its HZ. 

LTT~1445Ab is the nearest planet known to transit an M dwarf and is the most spectroscopically-accessible, terrestrial planet detected to-date.  LTT~1445Ab is a prime target for atmospheric studies due to its large transit depth ($\delta$ $=$ 0.2\%) and bright host star ($V_J =$ 11.22, $I_{KC} =$ 8.66, $K_s =$ 6.50 mag). It is also the second nearest known transiting planetary system to Earth. Currently, the nearest transiting planetary system is HD~219134bc \citep{Motalebi(2015),Gillon(2017)}, at a distance of 6.5 pc ($\pi$ $=$ 153.08$\pm$0.09 mas; \citealt{GaiaDR2(2018)}); but while the host star is brighter than \star ~($K_s$ $=$ 3.25 mag), the planetary transit depths are nearly an order of magnitude smaller ($\delta$ $=$ 0.036\% for HD~219134b). In contrast, the multi-planet system TRAPPIST-1 at 12.6 pc exhibits planetary transit depths that are, on average, a factor of three larger than that of \planet, but the host star is faint ($V_J =$ 18.75, $I_{KC} =$ 14.10, $K_s =$ 10.30 mag). 

There are many opportunities for follow-up studies of \planet. For instance, LTT~1445ABC is particularly favorable for ground-based observations to study the planet's atmosphere, as the blended BC pair may provide a valuable calibration source with the same spectral type as that of the primary star, although its utility as a comparison source may be limited if it is the source of the variation and flaring seen in \tess. \star ~is small enough and bright enough that we can collect enough photons in space for transmission spectroscopy. From \citet{Kempton(2018)}, we calculate a transmission spectroscopy metric (TSM) of 40 for \planet, where the TSM is the expected signal-to-noise ratio of transmission spectroscopy observations with {\it JWST}. This TSM is factors of four and three better than the TSMs of 9.15 and 13.7 for LHS~1140b and TRAPPIST-1f, respectively.

Based on the known occurrence rates of planets orbiting M dwarfs, it is unlikely that we will detect a small planet more favorable for atmospheric characterization than \planet. There are 55 mid-to-late M dwarf primaries closer than \star. The transit probability of \star ~is $1/30$. Assuming that all such stars have rocky planets, then we expect to find roughly one as amenable to follow-up study as \planet. However, \citet{Dressing(2015)} estimate the rate of occurrence of planets less than 1.5 \rearth ~to be 0.43 for orbital periods less than 10 days, in which case we expect that this is indeed the best one. A subtlety is that the \citet{Dressing(2015)} result pertains to early-type M dwarfs; the occurrence rates for mid-to-late M dwarfs may be higher, and thus we may find one or two more planets like \planet, if we are lucky.





\startlongtable
\begin{deluxetable*}{lcc}
\tablecaption{Median values and 68\% confidence interval for \planet \label{tab:pparams}}
\tablenum{8}
\tablehead{\colhead{~~~Parameter} & 
           \colhead{Units} & 
           \multicolumn{1}{c}{Values}}
\startdata
\smallskip\\\multicolumn{2}{l}{Stellar Parameters:}& A \smallskip\\
~~~~$M_*$\dotfill &Mass (\msun)\dotfill &$0.256\pm0.014$\\
~~~~$R_*$\dotfill &Radius (\rsun)\dotfill &$0.276^{+0.024}_{-0.019}$\\
~~~~$\rho_*$\dotfill &Density (g cm$^{-3}$)\dotfill &$17.3\pm3.9$\\
~~~~$\log{g}$\dotfill &Surface gravity (cgs)\dotfill &$4.967^{+0.061}_{-0.075}$\\
\smallskip\\\multicolumn{2}{l}{Planetary Parameters:}&b\smallskip\\
~~~~$T_0$\dotfill &Optimal conjunction Time (\bjdtdb)\dotfill &$2458423.42629\pm^{+0.00044}_{-0.00045}$\\
~~~~$P$\dotfill &Period (days)\dotfill &$5.35882^{+0.00030}_{-0.00031}$\\
~~~~$T_{14}$\dotfill &Total transit duration (days)\dotfill &$0.0574\pm0.0011$\\
~~~~$R_P/R_*$\dotfill &Radius of planet in stellar radii \dotfill &$0.0458^{+0.0012}_{-0.0011}$\\
~~~~$a/R_*$\dotfill &Semi-major axis in stellar radii \dotfill &$29.6^{+2.6}_{-2.5}$\\
~~~~$b$\dotfill &Transit Impact parameter \dotfill &$0.29^{+0.23}_{-0.20}$\\
~~~~$R_P$\dotfill &Radius (\rearth)\dotfill &$1.38^{+0.13}_{-0.12}$\\
~~~~$a$\dotfill &Semi-major axis (AU)\dotfill &$0.03807^{+0.00068}_{-0.00071}$\\
~~~~$i$\dotfill &Inclination (degrees)\dotfill &$89.40^{+0.41}_{-0.46}$\\
~~~~$e$\dotfill &Eccentricity \dotfill &$0.19^{+0.35}_{-0.14}$\\
~~~~$\omega_*$\dotfill &Argument of Periastron (degrees)\dotfill &$-139^{+120}_{-76}$\\
~~~~$T_{eq}$\dotfill &Equilibrium temperature (K)\dotfill &$433^{+28}_{-27}$\\
~~~~$K$\dotfill &RV semi-amplitude (m \pers)\dotfill &$2.1^{+1.6}_{-2.0}$\\
~~~~$K$\dotfill &RV semi-amplitude (m \pers)\dotfill &$<9.3$ [$3-\sigma$ upper limit]\\
~~~~$M_P$\dotfill &Mass (\mearth)\dotfill &$2.2^{+1.7}_{-2.1}$\\
~~~~$M_P$\dotfill &Mass (\mearth)\dotfill &$<8.4$ [$3-\sigma$ upper limit]\\
~~~~$\rho_{P}$\dotfill &Density (g cm$^{-3}$)\dotfill &$4.4^{+4.0}_{-4.2}$\\
~~~~$\rho_P$\dotfill &Density (g cm$^{-3}$)\dotfill &$<22$ [$3-\sigma$ upper limit]\\
~~~~$\log{g_P}$\dotfill &Surface gravity (cgs) \dotfill &$3.11^{+0.22}_{-0.34}$\\
~~~~$\log{g_P}$\dotfill &Surface gravity (cgs) \dotfill &$<3.7$ [$3-\sigma$ upper limit] \\
\smallskip\\\multicolumn{2}{l}{Wavelength Parameters:}&$TESS$\smallskip\\
~~~~$u_{1}$\dotfill &linear limb-darkening coeff \dotfill &$0.195\pm0.030$\\
~~~~$u_{2}$\dotfill &quadratic limb-darkening coeff \dotfill &$0.427\pm0.027$\\
~~~~$A_D$\dotfill &Dilution from neighboring stars$^{*}$ \dotfill &$0.000\pm0.013$\\
\smallskip\\\multicolumn{2}{l}{Telescope Parameters:}&HARPS\smallskip\\
~~~~$\gamma$\dotfill &Systemic RV (m \pers)\dotfill &$-5432.3\pm2.1$\\
~~~~$\dot{\gamma}$\dotfill &RV slope (m \pers day$^{-1}$)\dotfill &$-0.01275\pm-0.00053$\\
~~~~$\ddot{\gamma}$\dotfill &RV quadratic term (m \pers day$^{-2}$)\dotfill &$0.00000124\pm0.00000040$\\
~~~~$\sigma_J$\dotfill &RV Jitter (m \pers)\dotfill &$3.25^{+1.2}_{-0.80}$\\
\enddata
\tablecomments{$^{*}$fixed parameter}
\end{deluxetable*}


\vspace{5mm}
\begin{center}
\large
Acknowledgments
\end{center}
\normalsize

The authors thank the anonymous referee for their prompt response and for their comments and suggestions. We are indebted to Guillermo Torres for developing the M dwarf-specific relations from which we calculated the \tess ~magnitudes. We thank Mark Everett for his role in obtaining the DSSI speckle data. We are extremely grateful to Matthew Payne, Chelsea Huang, Joseph Rodriguez, Samuel Quinn,  and Jacob Bean for illuminating conversations and suggestions that helped improve the analysis and interpretation of the results presented here. 

The MEarth Team gratefully acknowledges funding from the David and Lucille Packard Fellowship for Science and Engineering (awarded to D.C.). This material is based upon work supported by the National Science Foundation under grants AST-0807690, AST-1109468, AST-1004488 (Alan T. Waterman Award), and AST-1616624. This work is made possible by a grant from the John Templeton Foundation. The opinions expressed in this publication are those of the authors and do not necessarily reflect the views of the John Templeton Foundation. Funding for the \tess ~mission is provided by NASA's Science Mission directorate. We acknowledge the use of public \tess ~Alert data from pipelines at the \tess ~Science Office and at the \tess ~Science Processing Operations Center. Resources supporting this work were provided by the NASA High-End Computing (HEC) Program through the NASA Advanced Supercomputing (NAS) Division at Ames Research Center for the production of the SPOC data products. HARPS observations were made with European Space Observatory (ESO) Telescopes at the La Silla Paranal Observatory under program IDs 072.C-0488, 183.C-0437, and 1102.C-0339. This research has made use of the Washington Double Star Catalog, maintained at the U.S. Naval Observatory. This work has made use of data from the European Space Agency (ESA) mission {\it Gaia} (\url{https://www.cosmos.esa.int/gaia}), processed by the {\it Gaia} Data Processing and Analysis Consortium (DPAC, \url{https://www.cosmos.esa.int/web/gaia/dpac/consortium}). Funding for the DPAC has been provided by national institutions, in particular the institutions participating in the {\it Gaia} Multilateral Agreement.


AAM and HD-L are supported by NSF Graduate Research Fellowship grants DGE1745303 and DGE1144152, respectively. NA-D acknowledges the support of FONDECYT project 3180063. EPH gratefully acknowledges funding from NSF grant AST-1517824. The RECONS team is indebted to long-term support from the NSF, most recently under grant AST-1715551. Work by JNW was partly supported by the Heising-Simons Foundation. Support for JKT and the acquistion of the speckle data was provided by NASA through Hubble Fellowship grant HST-HF2-51399.001 awarded by the Space Telescope Science Institute, which is operated by the Association of Universities for Research in Astronomy, Inc., for NASA, under contract NAS5-26555. XD is supported by the French National Research Agency in the framework of the Investissements d’Avenir program (ANR-15-IDEX-02), through the funding of the "Origin of Life" project of the Universit\'e Grenoble-Alpes. CDD acknowledges support from the \tess ~Guest Investigator Program through grant 80NSSC18K1583.

\vspace{5mm}
\facilities{TESS}, {MEarth}, {Gemini:North (DSSI)}, {FLWO:1.5m (TRES)}, {ESO:3.6m (HARPS)}, {CTIO:0.9m} 

\software{{\sc celerite} \citep{Foreman-Mackey(2017)}, {\sc exofastv2} \citep{Eastman(2013),Eastman(2017)}, {\sc exoplanet} \citep{exoplanet:exoplanet}, IDL, IRAF, {\sc PYMC3} \citep{exoplanet:pymc3}, {\sc python}, {\sc starry} \citep{exoplanet:luger18}}

This research made use of {\sc exoplanet} \citep{exoplanet:exoplanet} and its dependencies \citep{exoplanet:astropy13, exoplanet:astropy18, exoplanet:exoplanet, exoplanet:foremanmackey17, exoplanet:foremanmackey18, exoplanet:kipping13, exoplanet:luger18, exoplanet:pymc3, exoplanet:theano}.

\bibliographystyle{aasjournal}
\bibliography{masterref.bib}

\end{document}